\documentclass[11pt]{article}
\usepackage[utf8]{inputenc}
\usepackage[british]{babel}
\usepackage{datetime}

\usepackage{
	amssymb,
	amsmath,
	amsthm,
	mathabx,
	mathrsfs,
	stmaryrd,
	palatino,
	thm-restate
} 

\usepackage{
	hyperref,
	cleveref}

\usepackage{geometry}
\usepackage[all]{xy}
\usepackage{proof}
\usepackage{tikz} 
\usetikzlibrary{
  automata,
  positioning,
  arrows,
  shapes.geometric,
  calc
}

\tikzstyle{automaton} = [
  ->,
  > = stealth,
  node distance = 2cm,
  every state/.style = {
    thick,
    fill=black!0.0,
    draw=black!0.0
  }
]
\usepackage{mathpartir}

\usepackage[
style=numeric,
sorting=ynt
]{biblatex}
\newcommand{\resp}{resp.\  }

\newcommand{\ie}{i.e.\ }
\newcommand{\eg}{e.g.\ }

\newcommand{\catfont}[1]{\mathsf{#1}}
\newcommand{\Set}{\catfont{Set}}
\newcommand{\Alg}[1][F]{\catfont{Alg}\hspace{-1pt}\left(#1\right)}
\newcommand{\Coalg}[1][T]{\catfont{CoAlg}\hspace{-1pt}\left(#1\right)}
\newcommand{\Cov}{\catfont{Cov}}
\newcommand{\op}{^{\mathrm{op}}}
\newcommand{\BA}{\mathsf{BA}}
\newcommand{\DL}{\mathsf{DL}}
\newcommand{\id}{\mathrm{id}}
\newcommand{\mono}{\rightarrowtail}
\newcommand{\epi}{\twoheadrightarrow}

\newcommand{\coker}{\mathrm{coker}}

\newcommand{\Id}{\mathrm{Id}}
\newcommand{\Uf}{\mathcal{U}}
\newcommand{\Pow}{\mathcal{P}}
\newcommand{\At}{\mathsf{At}}
\newcommand{\free}{\mathsf{F}}
\newcommand{\forg}{\mathsf{U}}

\newcommand{\inv}{^{-1}}
\newcommand{\defeq}{\triangleq}
\newcommand{\im}{\mathrm{im}}
\newcommand{\pow}{{\mathcal P}}
\newcommand{\N}{\mathbb N}
\newcommand{\sem}[1]{\left\llbracket#1\right\rrbracket}
\newcommand{\coperp}{\hspace{3pt}\top\hspace{3pt}}

\newcommand{\trans}[1]{\mathbin{\raisebox{-1pt}{\(\xrightarrow{#1}\)}}}
\newcommand{\bisim}{\mathbin{\raisebox{0.5pt}{\(\underline{\leftrightarrow}\)}}}

\newtheoremstyle{thmstyle}
{1em} 
{1em} 
{\itshape} 
{} 
{\bfseries} 
{} 
{3pt} 
{\thmname{#1}\thmnumber{ #2.}\thmnote{ (#3)\\}}  

\newtheoremstyle{defstyle}
{1em} 
{1em} 
{} 
{} 
{\bfseries} 
{} 
{3pt} 
{\thmname{#1}\thmnumber{ #2.}\thmnote{ (#3)}}  

\theoremstyle{thmstyle}	
\newtheorem 	 	{theorem}         	{Theorem}[section]

\newtheorem 	 	{proposition}     	[theorem]{Proposition}
\newtheorem	   		{lemma}           	[theorem]{Lemma}

\theoremstyle{defstyle} 

\newtheorem 	  	{example}       	[theorem]{Example}

\title{How to write a coequation}
\author{Fredrik Dahlqvist\footnote{Department of Computer Science, UCL. \url{https://fredrikdahlqvist.wordpress.com}}
	 \and 
	 Todd Schmid\footnote{Department of Computer Science, UCL. \url{http://www.homepages.ucl.ac.uk/\~ucabtws}}
}
\date{ }

\begin{document}

\maketitle

\begin{abstract}
There is a large amount of literature on the topic of covarieties, coequations and coequational specifications, dating back to the early seventies.
Nevertheless, coequations have not (yet) emerged as an everyday practical specification formalism for computer scientists. 
In this review paper, we argue that this is partly due to the multitude of syntaxes for \emph{writing down} coequations, which seems to have led to some confusion about what coequations are and what they are for. 
By surveying the literature, we identify four types of syntaxes: \emph{coequations-as-corelations}, \emph{coequations-as-predicates}, \emph{coequations-as-equations}, and \emph{coequations-as-modal-formulas}.  
We present each of these in a tutorial fashion, relate them to each other, and discuss their respective uses.
\end{abstract}


\section{Introduction}
\label{sec:intro}

Characterising algebras by the equations they satisfy is common practice.
Equations are simple to write down: they consist of a pair of terms, or elements of an initial algebra.
Equations are also simple to interpret: terms denote constructions, so an equation between two terms asserts that two constructions produce equivalent objects.
The terms-as-constructions interpretation of equations is prevalent in programming language theory.
A programming language is a syntax for denoting programs, so equations state equivalences between programs.
Their importance can be seen in \(\lambda\)-calculus~\cite{barendregt1985lambda,salibragoldblatt1999finite,salibra2000algebraic,manzonettosalibra2008universal},  process algebra~\cite{fokkink2013introduction}, Kleene algebra~\cite{salomaa1966two,kozen1991completeness,conway2012regular} and its extensions~\cite{DBLP:conf/csl/KozenS96,fosterkozenmilanosilvathompson2015netkat,kappebrunetsilvazanasi2017concurrent,jipsen2014concurrent,smolkafosterhsukappekozensilva2019gkat,schmidkappekozensilva2021gkat}, and related areas~\cite{wechler1992universal,plotkins2001}.

A common thread, running through the many examples of equational reasoning in computer science, is that equations can be used to state \emph{behavioural} equivalences between programs.
For the purposes of this review, behaviours are what are obtained from dualizing, in the category theoretic sense, the concept of term. 
That is, a behaviour is an element of a final \emph{coalgebra}.
The dual study to algebra, coalgebra, constitutes a whole subfield of computer science dedicated to state-based dynamical systems, the sort of systems that exhibit behaviours~\cite{rutten1996universal,gumm2001products,jacobs2017introduction}.
Dualizing algebra not only takes terms to behaviours, but also equations to \emph{coequations}, the main focus of this article.

Broadly, a coequation is a constraint on the dynamics of a state-based system.
A coalgebra satisfies a coequation if its dynamics operate within the constraint.
This situation is familiar to those working in automata theory, since deterministic automata are examples of state-based dynamical systems. 
For a fixed alphabet \(A\), any set of languages \(\mathcal L \subseteq 2^{A^*}\) determines a coequation satisfied by those automata that only accept languages in \(\mathcal L\).
We call these coequations \emph{behavioural}, as they consist of a set of states in the final automaton \(2^{A^*} \to 2 \times (2^{A^*})^A\).
Not all coequations are behavioural:
following \cite{venema2007algebrasandcoalgebras,cirstea2011modal} in viewing Kripke frames as coalgebraic dynamical systems for the powerset functor, modal formulas provide illustrative examples of nonbehavioural coequations.
For example, reflexivity is a modally definable constraint on the dynamics of frames (witnessed by the modal formula \(\Box p \to p\)), despite the following two Kripke frames being behaviourally indistinguishable.
\[
\begin{tikzpicture}
	\node (0) {\(\bullet\)};
	\draw (0) edge[loop left, ->] (0);
\end{tikzpicture}
\qquad\qquad
\begin{tikzpicture}
	\node (0) {\(\bullet\)};
	\node[right of=0] (1) {\(\bullet\)};
	\node[right of=1] (2) {\(\bullet\)};
	\node[right of=2] (3) {\(\cdots\)};

	\path (0) edge[->] (1);
	\path (1) edge[->] (2);
	\path (2) edge[->] (3);
\end{tikzpicture}
\]\vspace{-2em}\\
Rather,  reflexivity is a coequation which requires two \emph{colours} to be stated, $p$ and $\neg p$. 
The concept of colour (or label) is key to moving beyond purely behavioural specifications, and can be understood as the formal dual to the notion of variable in algebra. 
Just like the commutativity of a binary operation requires two variables to be stated, the reflexivity of a Kripke frame requires two colours.

Coequations have arguably not seen much use by computer scientists, in spite of a large body of theoretical results.
We postulate that one of the main reasons for this is that there is no one universally accepted way to \emph{write down a coequation}.
It is not hard to see why: while an equation relates two, finite, tree-structures which can be written-down unambiguously in one dimension with the use of brackets, there is no universal syntax for describing \emph{constraints} on structures which are often \emph{inherently infinitary}. 
In fact, a variety of syntaxes for writing down coequations have been proposed in the literature, leading to a certain ambiguity surrounding the term coequation, especially since some of them are less expressive than others. 
This being said, it could equally be argued that coequations are used extensively, if unknowingly, by computer scientists, in the shape of modal logics \cite{blackburn2006handbook}.
For example, 
languages like Linear Temporal Logic \cite{vardi1986automata,goldblatt1987logics,gerth1995simple} and Computation Tree Logic \cite{emerson1985decision,clarke1986automatic} are efficient syntaxes for specifying coequations. 

The purpose of this paper is three-fold.  The first is to survey and organise the literature on coequations. 
The second is to act as a tutorial on coequations and coequational specification.
We assume basic knowledge of category theory 
 and focus on $\Set$-coalgebras. 
Finally, we aim to present a systematic account of what the various notions of coequations are, how they are related to one another, and what role they have to play in theoretical computer science.  

The paper is structured as follows. 
We start with a review of the literature on coequations in \cref{sec:lit_review}, highlighting a number of formalisms for defining and specifying coequations. 
We group these approaches into four paradigms, which we examine in detail. 
First, in \cref{sec:corelations}, we present a notion of coequation which dualizes exactly the notion of equation in Universal Algebra, and which we call \emph{coequation-as-corelation}. 
Second, we present the view that a coequation is a predicate on a cofree coalgebra. 
We call this notion \emph{coequation-as-predicate} and discuss it in detail in \cref{sec:predicates}. 
Third, we discuss \emph{coequations-as-equations} in \cref{sec:coeq-eq} and relate them to coequations-as-corelations. 
The last paradigm we explore is that of coequations as modal formulas in \cref{sec:modal}.  
Finally, we conclude in \cref{sec:prospects} with some thoughts on the uses of each formalism and some recent appearances  of coequations in computer science.


\section{A brief history of coequations}\label{sec:lit_review}
As far as we are aware, the first mention of the word `coequation'-- or more precisely `coequational' -- dates back a series of papers by Davis \cite{davis1970universal,davis1972multivalued,davis1972cotripleable,davis1983combinatorial,davis1984combinatorial} starting in 1970 and focusing on finding examples of comonadic/cotripleable categories.
The earliest work that deals with covarieties of coalgebras as we now understand them,  seems to be the 1985 paper \cite{marvan1985covarieties}, which presents a category-theoretic account of a dual to Birkhoff's HSP theorem. 
This work focuses on coalgebras for polynomial functors on the category $\Set$. 
It describes an unusual approach to dualising Birkhoff's HSP theorem which reduces the problem to the ordinary version of the theorem by turning every coalgebra $X\to FX$ into an algebra $2^{FX}\to 2^X$, and by introducing an infinitary equational logic extending that of complete atomic boolean algebras to define varieties of such algebras.  
The idea of establishing a bridge between coequations and equations was explored again in \cite{ballesterbolinchescosmellopezrutten2015dual} and  \cite{salamanca2016dualityofequations} where conditions for a full duality between equations and coequations  are given. 

A few years after \cite{marvan1985covarieties}, work on the notion of \emph{terminal coalgebra} in \cite{aczel1988nwfs,aczel1989final,barr1993terminal} laid the ground for \cite{hensel1994defining, reichel1995approach, jacobs1995mongruences}, which proposed coalgebras as a semantic framework to formalise behaviours in object-oriented programming and infinite data structures such as streams and trees.  
From the onset, the aim of this line of research was to syntactically specify classes of behaviours and, although the terms `covariety' and `coequation' do not appear in \textit{op.cit.}, a lot of the questions which we will explore in this paper can already be found in Hensel and Reichel's \cite{hensel1994defining} and Jacobs' \cite{jacobs1995mongruences}. 
Both approaches propose \emph{equational specifications} of coalgebras for polynomial functors, based on the signature of the functor.  
For example, the equation $\mathsf{head(tail(x))}=\mathsf{head(x)}$, where the `destructor' signature $\mathsf{head},\mathsf{tail}$ can be read off the functor $F(X)=X\times A$, characterises constant streams in the terminal $F$-coalgebra $A^\omega$.  
Such equations are called \emph{state equations} in \cite{hensel1994defining} since they must hold at every state, and \cite{jacobs1995mongruences} gives a concrete construction of the class of behaviours satisfying such an equation via the notion of `mongruence' (a terminology which mercifully has not caught on).  
This kind of coequational specification via equations, which we refer to as \emph{coequations-as-equations}, is also used in C{\^\i}rstea's 1999 \cite{cirstea1999coequational}, which is the first full paper to use the term `coequation' in the sense we understand today.  
Ro{\c{s}}u's \cite{rocsu2001equational} from 2001 follows the same approach of \emph{coequations-as-equations}. 
While intuitive, this way of `writing coequations' is limited to endofunctors of a specific shape.
Another, more powerful, way of writing coequations-as-equations was developed by Kurz and Rosick{\`y} in \cite{kurz2005operations} using an equality relation between terms built from a signature, but with different notions of (co)operation and term. 
In this framework, every covariety over $\Set$ can be presented in a coequation-as-equation format.

Jacobs' \cite{jacobs1995mongruences} pioneered this specification format but also asked the following questions, which motivated a lot of the subsequent research on the topic: 1) Can a sound and complete logic to reason about coalgebras be devised? 2) Can a version of Birkhoff's theorems be proved for suitable classes of coalgebras?
The first step towards answering these question was taken by Rutten's influential 1996 technical report \cite{rutten1996universal,rutten2000universal}, which introduces the notion of `colours' of a cofree coalgebra,  the concept of \emph{covariety}, and the first of many dual versions of Birkhoff's HSP theorem. 
The term `coequation' does not appear in \cite{rutten1996universal,rutten2000universal}, but it is worth noting that a concrete specification of a covariety is given by a \emph{subcoalgebra of a cofree coalgebra}, in contrast to the equational presentation of \cite{hensel1994defining,jacobs1995mongruences,cirstea1999coequational,rocsu2001equational}.  
We refer to this type of coequational specification as \emph{coequations-as-predicates}. 

An important moment in the history of coequations was the first Workshop on Coalgebraic Methods in Computer Science (CMCS),  organised by Jacobs,  Moss,  Reichel and Rutten and held in Lisbon in 1998.  
Several papers on covarieties were presented \cite{gumm1998covarieties,kurz1998specifying, rocsu1998birkhoff}, and the next few years saw an explosion of research in this area.  
In retrospect, CMCS 1998 provided much of the momentum behind the subsequent blossoming of this new field of research.

In their 1998 CMCS paper \cite{gumm1998covarieties}, Gumm and Schr\"{o}der  pick up the study of covarieties defined via a subcoalgebra in \cite{rutten1996universal}. 
They isolate precisely which subcoalgebras define a covariety and describe the closure properties of covarieties closed under bisimulation, which they call \emph{complete covarieties}. 
It was subsequently shown in \cite{awodey2000coalegebraic, hughes2001study} that these covarieties, more aptly called \emph{behavioural covarieties}, are precisely those which can be described by a coequation-as-predicate over one colour, that is to say by a subcoalgebra of the terminal coalgebra.  
Coequations as subcoalgebras of a cofree coalgebras (or more abstractly as regular monomorphisms with cofree codomain) and the covarieties they define are also discussed in detail in \cite{adamek2001varieties,adamek2003varieties} where a dual to Birkhoff's HSP theorem is given.

Hughes' 2001 thesis \cite{hughes2001study} presents a very detailed abstract account of the \emph{coequations-as-predicates} perspective.  
A coequation is no longer required to be defined by a sub\emph{coalgebra}, but can simply be a sub\emph{set} of a cofree coalgebra \cite[\S 3.6.3]{hughes2001study}.  
Closure operators defined and studied in \cite{hughes2001study,hughes2001modal,jacobs2002temporal} connect these two flavours of the \emph{coequations-as-predicates} paradigm, by constructing the (invariant) sub\emph{coalgebra} generated by a sub\emph{set} of behaviours.  
More abstractly, \cite{hughes2001study} also considers a coequation as a subcoalgebra of a regular injective coalgebra (\eg a cofree coalgebra). 
This additional abstraction dualizes the description of equations as quotients of regular projective algebras due to \cite{banaschewski1975subcategories}, but introduces subtle differences on the closure properties of covarieties which are discussed by Goldblatt in \cite{goldblatt2005comonadic,clouston2005covarieties}. 
In this paper, we only consider subcoalgebras of cofree coalgebras or subsets of their carriers.

A third version of the \emph{coequations-as-predicates} paradigm was proposed by Gumm in \cite{gumm2001equational}, where a coequation is a single pattern (\ie element of a cofree coalgebra), but a pattern that \emph{must be avoided}. 
In other words, this is a \emph{coequation-as-predicate} defined by the complement of a singleton, \ie understood as a \emph{pattern avoidance constraint}.  
This fruitful idea was the source of many interesting examples in \cite{adamek2003varieties}, and the basis for coequational logics in \cite{adamek2005logic,schwencke2008coequational,schwencke2010coequational}.  
These logics are based on two observations which can already be found in \cite{gumm2001equational}, namely that if there exists a state $x$ witnessing a pattern $f$,  then any `successor pattern' of $f$ must be witnessed by some state $y$ (namely one of the successors of $x$). 
Similarly, if there exists a state $x$ witnessing a pattern $f$,  there must exist a recolouring/relabelling of this state which witnesses a similar relabelling of the pattern $f$.  
Adamek \cite{adamek2005logic} shows that these two observations are enough to define a sound and complete coequational system in which new coequations (avoidance patterns) can be deduced from known ones. 
This logic is most natural for polynomial functors, but can also be made to work for very large class of accessible functors through the notion of functor presentation \cite{adamek2005logic,schwencke2008coequational,schwencke2010coequational}.

Finally, we mention generalisations of \emph{coequations-as-predicates} to the case where cofree coalgebras do not exist.  
Ad\'amek and Porst generalise coequations-as-predicates by considering regular monomorphisms into \emph{any} element of the cofree coalgebra chain, whether it stabilises or not \cite{adamek2003varieties}.  
Kurz and Rosick{\`y} in \cite{kurz2005operations} describe the notion of \emph{implicit operations} which permits an equivalent notion of coequation-as-predicate for functors which have no cofree coalgebras.  
Finally, Ad\'amek describes a comprehensive solution to this problem by considering generalised coequations-as-predicates as subchains of the entire cofree-coalgebra chain in \cite{adamek2005birkhoff}.

A related but different notion of coequation was proposed in 2000 by Wolter \cite{wolter2000corelations} and Kurz \cite{kurz2000phd}, systematically dualizing the picture from categorical Universal Algebra.
Since a set of equations can be understood as a relation between terms in a free algebra (categorically a span), Wolter proposes that a coequation should be seen as a \emph{corelation} on the carrier of a cofree coalgebra over some set of colours (categorically, a cospan).  
Similarly, Kurz proposes to consider a \emph{cocongruence} on the cofree coalgebra. 
As in the \emph{coequation-as-predicate} paradigm, these two approaches reflect the fact that one may consider a coequation as a structure on the \emph{carrier} of a cofree coalgebra (a corelation), or on the cofree coalgebra itself (a cocongruence).
We will refer to this approach as \emph{coequations-as-corelations}.  This approach neatly dualizes the well-known theory of equations, but the notion of corelation is not very intuitive, as pointed out by Hughes \cite{hughes2001study}. Nevertheless, we hope to provide some intuition in \cref{sec:corelations} and \cref{sec:coeq-eq}.

At the same period Kurz also proposed \emph{modal logic} as a language for specifying covarieties \cite{kurz1998co,kurz2000phd,kurz2001modal,kurz2001specifying}.  
This is our last paradigm for coequations: \emph{coequations-as-modal-formulas}. 
Following our discussion in the introduction, we know that for finitely branching Kripke frames there exists a \emph{final} Kripke frame coloured by (sets of) propositional variables, and any modal formula $\phi$ selects the states in this Kripke frame in which $\phi$ is valid, \ie defines a predicate on a cofree coalgebra. 
A modal formula can thus be seen as a syntax for \emph{coequations-as-predicates}. 
However, these are very particular predicates: they have a simple and intuitive syntax,  access to a countable set of colours (the propositional variables), and to 1-step ahead colours (through modalities).  
This idea can easily be extended to modal logics for polynomial functors \cite{kurz2001modal}. 
However,  the idea of modal logic as a specification language for covarieties found in \cite{kurz1998co,kurz2000phd,kurz2001modal,kurz2001specifying} was in some way too prescient: the appropriate extension of modal logic -- \emph{coalgebraic modal logic} -- was still in its infancy.
Moss' logic \cite{moss1999coalgebraic} had only just been published, and neither the predicate lifting formalism of Pattinson \cite{pattinson2003coalgebraic} nor the abstract formalism of Kupke, Kurz, and Pattinson \cite{kupke2004algebraic,kupke2005ultrafilter,jacobs2010exemplaric} had been developed. 
As a consequence,  Kurz's insight of \emph{coequations-as-modal-formulas} was only worked out for standard modal logics \cite{kurz2001modal}. His abstract notion of \emph{modal predicate} \cite{kurz2000phd} -- where the term `modal' is meant as `invariant under bisimulation' -- is not based on a particular syntax, but is defined as a monomorphism into a cofree coalgebra, \ie as a coequation-as-predicate. 

\section{Coequations-as-corelations}\label{sec:corelations}
\emph{Coequations-as-corelations} is the notion of coequation which most faithfully dualizes the notion of equation from Universal Algebra. 
It is not the simplest approach, but it exposes the underlying machinery in its entirety.  Syntactically, it is a formalism that resembles equations because it uses a \emph{pairs of expressions}.  
However,  whilst an equation between a pair of expressions \emph{forces} an equality to be witnessed via a quotient, a \emph{coequation-as-corelation} involves a pair of expressions which \emph{selects} an `existing' equality via an equaliser. 
Much of the material in this section can be found in \cite{wolter2000corelations,kurz2000phd,awodey2000coalegebraic,hughes2001modal,hughes2001study,dahlqvist2015phd}.
We present the classical picture from Universal Algebra in \cref{subsec:alg},  and then dualize it in \cref{subsec:coalg}.  

\subsection{Equations, relations and varieties of algebras.}\label{subsec:alg}

We will not go to the level of generality of \cite{awodey2000coalegebraic,hughes2001study,hughes2001modal} but instead focus on algebras for $\Set$-endofunctors.  
Let $T:\Set\to\Set$ be an endofunctor and let $\Alg[T]$ denote the category of $T$-algebras and $T$-algebra morphisms. 
There exists an obvious forgetful functor $U_T: \Alg[T]\to \Set$ which keeps the carrier and forgets the algebraic structure. 
A functor $T$ is called a \emph{varietor} \cite{adamek1990automata} if this functor has a left-adjoint $F_T:\Set\to \Alg[T]$ which builds free $T$-algebras over any given set of variables.  
We will drop the subscripts and simply write $F\dashv U$ if this causes no ambiguity.  
It follows from the adjunction that any map $h: X\to U(A,\alpha)$ can be freely extended to a $T$-algebra morphism $\hat{h}: FX\to(A,\alpha)$ explicitly constructed as $\hat{h}\defeq\varepsilon^T_{(A,\alpha)}\circ Fh$, where $\varepsilon^T$ is the counit of the adjunction.

For any varietor $T$ we define a
 \emph{set of $T$-equations over a set of variables $X$} is a pair of arrows $e_1,e_2: E\rightrightarrows UF X$.  A set of equations is thus represented as a \emph{span}, the categorical embodiment of the notion of relation.

A $T$-algebra $(A,\alpha)$ \emph{satisfies a set of equations $e_1,e_2: E\rightrightarrows UF X$}  if for all \emph{valuations} $v: X\to U(A,\alpha)$,  $U\hat{v}\circ e_1=U\hat{v}\circ e_2$, \ie if the map $\hat{v}$ which recursively computes the interpretation in $(A,\alpha)$ of formal terms from $FX$, returns the same output for the left- and right-hand-side of each equation in $E$. 
This can be rephrased as a universal property in $\Alg[T]$ by saying that $(A,\alpha)$ satisfies a set of equations $e_1,e_2: E\rightrightarrows UF X$ if any $T$-algebra morphism $f: FX\to(A,\alpha)$ factors uniquely through the coequalizer \(q\) of $\hat{e}_1,\hat{e}_2$.
\begin{align}
		\xymatrix@C=10ex@R=4ex
		{
			F E \ar[r]<3pt>^-{\hat{e}_1}  \ar[r]<-1pt>_-{\hat{e}_2}  & F X \ar@{->>}[r]^{q} \ar[d]^-{f}& (Q,\nu)\ar@{-->}[dl]\\
			& (A,\alpha)
		}
		\label{diag:equation}
\end{align}
Since any morphism $f: FX\to(A,\alpha)$ is of the shape $f=\hat{v}$ for some $v: X\to U(A,\alpha)$, we recover the standard notion of equation satisfaction.  
An object $(A,\alpha)$  with the universal property in \eqref{diag:equation} is said to be \emph{orthogonal} to $q: FX\epi (Q,\nu)$, written $q\perp (M,\alpha)$.

The \emph{variety of $T$-algebras defined by the set of $T$-equations} $e_1,e_2:E\rightrightarrows UF X$ is defined as the class of all $T$-algebras which are orthogonal to the coequalizer of the adjoint morphisms $\hat{e}_1,\hat{e}_2: FE\rightrightarrows FX$, notation $q^\perp$. 
Equivalently, a variety of $T$-algebras is a class of $T$-algebras orthogonal to a regular epi $q: F X\epi Q$,\footnote{By taking $U\ker(q)\rightrightarrows UFX$ as the set of equation and using the fact that $U$ is monadic \cite[20.56]{adamek2004abstract}, it can be shown that we recover the quotient $q$ as the coequalizer of the lifted equations.} a definition which dates back to \cite{banaschewski1975subcategories}.  With this terminology in place we state Birkhoff's famous HSP theorem.

\begin{theorem}[\cite{birkhoff1935structure,sankappanavar1981course,hughes2001study,adamek2001varieties,adamek2003varieties}]\label{thm:HSP}
Let $T:\Set\to\Set$ be a varietor. A class of $T$-algebras is a variety iff it is closed under Homomorphic images (H), Subalgebras (S), and Products (P).
\end{theorem} 

\begin{example}\label{ex:monoids}
Recall that a monoid is a set $M$ equipped with a binary operation $*:M\times M\to M$ and a constant $e\in M$ satisfying the three \emph{equations}: 
$x*(y*z) = (x*y)*z$, $e*x = x$, and $x*e=x$.
Every monoid is an algebra for the functor $\Sigma M = M\times M + 1$, and the functor $\Sigma$ is a varietor: $F_\Sigma$ builds the set of all formal terms constructed from the signature and a set of variables (\eg $\{x,y,z\}$), and equips it with the trivial $\Sigma$-algebra structure taking the unit to be the \emph{term} $e$, and the product of two terms $s,t$ to be the \emph{term} $s * t$. 
The equations of the theory of monoids can be described as the pair of maps
\(
	e_1, e_2: 3\rightrightarrows UF \{x,y,z\}
\),
where $3\defeq\{0,1,2\}$ and for $0\leq i\leq 2$, $e_1(i)$ (\resp $e_2(i)$) picks the left-hand-side (\resp right-hand-side) of the $i^{th}$ equation above.  A monoid is a $\Sigma$-algebra in the variety defined by the coequalizer of the adjoint morphisms $\hat{e}_1,\hat{e}_2: F3\rightrightarrows F\{x,y,z\}$ which homomorphically sends formal terms on $3$ to terms on $\{x,y,z\}$, for example $\hat{e}_1(1*2)=(e*x)*(x*e)$. The relation defined by the span $U\hat{e}_1,U\hat{e}_2: UF3\rightrightarrows UF\{x,y,z\}$ is thus closed under the rule
\[
\hspace{-1ex}
\infer[\textbf{p-cong }]{s_1*s_2 = t_1*t_2}{s_1=t_1\quad s_2=t_2}
\]
For example, $(x*e)*(e*x)=x*x$ is an equation belonging to this relation. 
Such a relation on terms is called a \emph{pre-congruence} in \cite{hughes2001study}. 
\end{example}

The rule \textbf{p-cong} generalizes easily to all polynomial functors, but it is not obvious how it should be adapted to the general case.
Therefore, we simply say that the relation defined by a span on $UFX$ is a \emph{pre-congruence} if it is of the shape $U\hat{e}_1, U\hat{e}_2: UFE\rightrightarrows UFX$.

It is important to note that the quotient $(Q,\nu)$ in \eqref{diag:equation} is \emph{not} a member of the variety. 
In \cref{ex:monoids}, $y*e$ and $y$ belong to different equivalence classes in $Q$ since the quotient $q$ only needs to identify $x*e$ and $x$. 
The interpretations of $y*e$ and $y$ are equal in all objects belonging to the variety of monoids because of the universal quantification over the morphism $f$ in \eqref{diag:equation} which takes care of all substitutions. 
In order to build a quotient that \emph{does} belong to the variety we need more equations than those in the set $UFE$.  
It is well known that equational reasoning also adheres to the following rules:
\[
\infer[\textbf{ref}]{t=t}{}
\qquad\qquad
\infer[\textbf{sym}]{s=t}{t=s}
\qquad\qquad
\infer[\textbf{trans}]{s=u}{s=t\quad t=u}
\]
A relation on $UFX$ which is closed under \textbf{ref}, \textbf{sym}, and \textbf{trans} is called an \emph{equivalence relation}, and a pre-congruence which is also an equivalence relation is called a \emph{congruence}. 
It is easy to turn the relation defined by \eqref{diag:equation} into a congruence by taking the \emph{kernel pair} of the coequalizer $q$, \ie by moving to the exact sequence $\ker(q)\rightrightarrows FX\epi(Q,\nu)$. 
Since any coequalizer is also the coequalizer of its kernel pair, $q$ remains the coequalizer, and thus the variety it defines remains the same.
We have now increased the collection of derivable equations. 
Following \cref{ex:monoids}, the congruence $\ker(q)\rightrightarrows F\{x,y,z\}$ contains the equation $e*x=x*e$, for example, which requires \textbf{sym} and \textbf{trans} to derive. 

As the reader will have guessed, we need to add substitution instances.  
Starting from the pre-congruence of \eqref{diag:equation}, this can be done categorically~\cite[\S 1.4]{dahlqvist2015phd}  by considering the coequalizer
\begin{align}
\xymatrix@C=12ex
{
\coprod\limits_{v\in V} F E \ar[r]<3pt>^-{\left[\hat{v}\circ\hat{e}_1\right]_{v\in V}}  \ar[r]<-1pt>_-{\left[\hat{v}\circ\hat{e}_2\right]_{v\in V}}  & F X \ar@{->>}[r]^{q'} & (Q',\nu')
}
\label{diag:stable}
\end{align}
where $V$ is the set of all substitutions $V=\{v: X\to UFX\}$. It is not difficult to see that $(Q',\nu')$ now \emph{does} belong to the variety defined by $q': FX\epi (Q',\nu')$, \ie $q'\perp (Q',\nu')$.

Since $F$ is a left-adjoint, $\coprod_{v\in V} F E\simeq F(\coprod_{v\in V} E)$, \ie \eqref{diag:stable} involves the free $T$-algebra generated by all substitution instances of the axioms $UFE$. The relation defined by the span $U\coprod_{v\in V} FE \rightrightarrows UFX$ is a pre-congruence closed under the substitution rule
\[
\infer[\textbf{subst}]{\hat{v}(s) = \hat{v}(t)}{s = t\qquad v\in V}
\]
Applying \eqref{diag:stable} to \cref{ex:monoids}, we get that $y*e=y$ now belongs to the stock of equations. 
In fact, it appears several times, since any substitution mapping $x$ to $y$ will produce it.

Following \cite{hughes2001study} we say that a set of equations $e_1,e_2: E\rightrightarrows UFX$ is \emph{stable} if it is closed under substitutions in the sense that for any $v\in V$ there exists a (necessarily unique) map $\tilde{v}: E\to E$ such that $e_i\circ \tilde{v}=U\hat{v}\circ e_i, i=1,2$.  Taking the kernel pair
\begin{align}
\xymatrix@C=12ex
{
\ker(q') \ar[r]<3pt> \ar[r]<-1pt> & F X \ar@{->>}[r]^{q'} & (Q',\nu')
}
\label{diag:stable2}
\end{align}
constructs a stable set of equations $U\ker(q')\rightrightarrows UFX$  by construction \cite[\S 1.4]{dahlqvist2015phd}.

We have described three categorical constructions -- lifting the equations,  closing under \eqref{diag:stable}, and taking the kernel pair of the coequalizer  \eqref{diag:stable2} -- which, combined, turn a \emph{set} of equations $e_1,e_2: E\rightrightarrows UFX$ into an exact sequence $\ker(q')\rightrightarrows FX \epi (Q',\nu')$ which defines a \emph{stable congruence} $U\ker{q'}\rightrightarrows UFX$.  
We do not know if this purely categorical procedure produces the \emph{smallest} stable congruence, and we do not know if the order in which the three steps are carried out matters.  
These questions are also raised in \cite[\S 8]{hughes2001modal}, and as far as we could see, no simple categorical argument 
can answer them.
What is clear however, is that this construction defines a quotient of the free $T$-algebra which belongs to the variety it defines.

\begin{theorem}[\cite{hughes2001study} Thm 3.5.3]\label{thm:stablecongruence}
Let $T$ be a varietor, let $e_1,e_2:E\rightrightarrows UFX$ be a stable set of $T$-equations over $X$, and consider the coequalizer
\[
\xymatrix
{
FE\ar@<3pt>^{\hat{e}_1}[r]\ar@<-1pt>[r]_{\hat{e}_2} & FX\ar@{->>}[r]^-{q} & (Q,\nu).
}
\]
Then $q\perp (Q,\nu)$. Conversely, if $q\perp(Q,\nu)$, then $\ker(q)$ is stable.
\end{theorem}

We finish by stating Birkhoff's completeness theorem for equational reasoning.  
Given a collection $\mathbb{V}$ of $T$-algebras (e.g. a variety), define $\mathrm{Eq}(\mathbb{V})$ as the set of equations satisfied by every $T$-algebra in $\mathbb{V}$, \ie
\[
\mathrm{Eq}(\mathbb{V})=\{e_1,e_2: 1\rightrightarrows UFX\mid \forall (M,\alpha)\in \mathbb{V}, \forall v: X\to M, U\hat{v}\circ e_1=U\hat{v}\circ e_2\}.
\]

\begin{theorem}[Birkhoff's completeness theorem,  \eg \cite{birkhoff1935structure,sankappanavar1981course,hughes2001study}]
Let $\Sigma$ be a polynomial functor, and let $E\rightrightarrows U_\Sigma F_\Sigma X$ be a set of equations.  Then $E=\mathrm{Eq}(\mathbb{V})$ for some variety $\mathbb{V}$ iff $E$ is closed under \textbf{p-cong}, \textbf{subst}, \textbf{ref}, \textbf{sym} and \textbf{trans} .
\end{theorem}

\subsection{Coequations, corelations and covarieties of coalgebras}\label{subsec:coalg}

Next, we dualize the concepts developed in \cref{subsec:alg}.
We denote by $\Coalg$ the category of $T$-coalgebras and $T$-coalgebra homomorphisms. 
A functor $T:\Set\to\Set$ is called a \emph{covarietor} \cite{adamek2001varieties} if the forgetful functor $U_T: \Coalg\to\Set$ has a \emph{right} adjoint $C_T: \Set\to\Coalg$, called the \emph{cofree functor}.  For any set $X$, the coalgebra $C_T X$ is called 
the \emph{cofree coalgebra over the set of colours $X$}, or the \emph{cofree coalgebra in $X$ colours}.  
We omit subscripts if there is no risk of confusion. 
Intuitively, $C_TX$ is the collection of \emph{$X$-pattern}s, $T$-processes (histories of states in a $T$-transition system) whose states are labelled by elements of $X$.

Given a covarietor $T$ we dualize the notion of equation by defining a \emph{$T$-coequation in $X$ colours}  \cite{wolter2000corelations} to be a \emph{cospan}
\(
c_1,c_2: UCX\rightrightarrows 2
\).
The role of $2\defeq\{0,1\}$ is dual to the role of \(1\) in the definition of a $T$-equation, since \(2\) is a cogenerator in $\Set$, whilst \(1\) is a generator.  
Following \cite{wolter2000corelations}, we define a \emph{$T$-coequational specification $S$} as a pair of maps $c_1,c_2: UCX\rightrightarrows S$. 
This concept dualizes the notion of a set of $T$-equations, and whilst a set of $T$-equations is equivalent to a relation on $UF X$, a coequational specification defines a \emph{corelation}, \ie a  map $UCX+UCX\epi S$.
One should think of a corelation on $UCX$ as two different classification schemes -- in the case of a coequation, two binary classification schemes, accepting or rejecting behaviours -- used to select the behaviours/patterns that they cannot distinguish.

The dual to the notions of valuation and interpretation/substitution are the notions of \emph{colouring} and \emph{recolouring map}.  
Given a $T$-coalgebra $(V,\gamma)$, a function $k:V\to Y$ is called a \emph{$Y$-colouring map}, as it labels the states of the coalgebra with the colours of $Y$.  
Given such a colouring map, we call its cofree extension $\hat{k}: (V,\gamma)\to CY$ a \emph{recolouring map}, $\hat{k}\defeq C_T k\circ \eta^T_{(V,\gamma)}$. 
Starting at a state $v\in V$, this map follows the history of the $T$-transition system $(V,\gamma)$, reads the colour(s) of the successor state(s) at each time step, and uses this information to construct the $T$-transition system of observed colours. 
The original $T$-history is typically infinite, and therefore so is the $T$-history of its colours. 
Thus, $\hat{k}$ is a map which typically processes an entire infinitary structure in one go.

Colouring maps of the shape $k: UCX\to Y$ are important to understanding coequations. 
Imagine an omniscient being that can examine the entire (possibly infinite) history of an $X$-pattern in $UCX$ and then classify it according to a rule of her choosing with a set of labels $Y$. 
This is a colouring map on $UCX$.  
In particular, for a colouring map $k: UCX\to X$, the omniscient being can examine the entire $X$-labelling history of a $T$-process and aggregate this information into a single $X$-label. 
Every cofree coalgebra comes with such a canonical colouring map $\varepsilon_X^T: UCX\to X$ provided by the counit $\varepsilon^T$ of the adjunction $U\dashv C$, which returns the colour of the initial state.  
Given a colouring map $k: UCX\to Y$, the associated recolouring map $\hat{k}:CX\to CY$ can be understood as the process by which our omniscient being can follow an entire $T$-process and, at each time-step, classify the \emph{remaining history} of the $T$-process according to the colouring map $k$. 
In this way the omniscient being can build the  \emph{labelling history} of the entire process in one single evaluation. 

The \emph{covariety defined by a $T$-coequational specification $c_1,c_2: UCX\rightrightarrows S$ over a set (of colours) $X$},  is the class of coalgebras $(V,\gamma)$ such that for every colouring map $k: V\to X$, there is a unique coalgebra morphism from $(V,\gamma)$ to the \emph{equaliser} $(H,\xi)$ of $\hat{c}_1,\hat{c}_2$ such that
\begin{align}
\xymatrix@C=12ex@R=4ex
{
(H,\xi)~ \ar@{>->}[r]^{m} & CX \ar@<3pt>[r]^{\hat{c}_1} \ar@<-1pt>[r]_{\hat{c}_2} & CS \\
& (V,\gamma) \ar[u]_-{\hat{k}} \ar@{-->}[ul]
}
\label{diag:covariety}
\end{align}
commutes.\footnote{For any $\Set$-endofunctor the category $\Coalg$ always has equalisers \cite[5.1]{gumm2001products}}
The coalgebra $(V,\gamma)$ is said to be co-orthogonal to the regular mono $m$, written $m \coperp (V,\gamma)$, and the covariety defined by \eqref{diag:covariety} can be described as the collection of coalgebras $m^{\top}$ which are co-orthogonal to $m$.
With this definition, we can state the dual to \cref{thm:HSP}:

\begin{theorem}[co-Birkhoff (HSC) theorem,  \eg  \cite{kurz2000phd,hughes2001study,adamek2001varieties,adamek2003varieties}]\label{thm:coBirkhoff}
Let $T:\Set\to\Set$ be a covarietor. A class of $T$-coalgebras is a covariety iff it is closed under Homomorphic images (H), Subcoalgebras (S) and Coproducts (C).
\end{theorem}

Dual to the case of varieties,  closure properties of the corelation $c_1,c_2: UCX\rightrightarrows S$ can ensure that we obtain an equaliser which belongs to the covariety.  
Most of the literature focuses on closure properties on the subobject side of \eqref{diag:covariety}, \eg the notion of mongruence \cite{jacobs1995mongruences}, the (modal) closure operators of \cite{hughes2001study,hughes2001modal}, and the notion of invariant subcoalgebra of \cite{gumm1998covarieties}. 
Since we dedicate \cref{sec:predicates} to this perspective, we follow \cite{wolter2000corelations,kurz2000phd} and focus on the quotient. 

Merging the nomenclatures of \cite{hughes2001study} and \cite{kurz2000phd}, we call the cospan $\hat{c}_1,\hat{c}_2: CX\rightrightarrows CS$ a \emph{pre-cocongruence}, the notion dual to a pre-congruence.  
Thus a pre-cocongruence is a corelation that has a coalgebra structure compatible with that of $CX$, \ie which is closed under taking successors.\footnote{This is what Kurz calls a cocongruence in \cite{kurz2000phd}.} 
Following \cite{wolter2000corelations}, we will say that a corelation $c_1,c_2: UCX\rightrightarrows S$  is \emph{coreflexive} if there exists a map $s: S\to UCX$ such that $s\circ [c_1,c_2]=[\id_{UCX}, \id_{UCX}]$, \ie if it is only allowed to identify a behaviour $(1,t)$ in the first component of the coproduct with a behaviour $(2,s)$ in the second if $t=s$. 
There is also a notion of cosymmetric, cotransitive, and of coequivalence corelation, but it turns out that in $\Set$ these are implied by being coreflexive \cite{wolter2000corelations}. 
Following our earlier definition of pre-congruence and congruence, we will say that a pre-cocongruence is a cocongruence if it is coreflexive. 
It is easy to turn any corelation into a coreflexive corelation, it suffices to consider the \emph{cokernel of its equaliser}.  
Finally,  dual to the notion of a stable set of equations, we will say that a corelation $c_1,c_2: UCX\rightrightarrows S$ is \emph{invariant} if for any colouring map $k: UCX\to X$ there exists a (necessarily unique) morphism $\tilde{k}: S\to S$ such that $c_i\circ \tilde{k}=U\hat{k}\circ c_i, i=1,2$. 
By dualizing \eqref{diag:stable}-\eqref{diag:stable2} we can turn any corelation into an invariant corelation by first considering the equalizer of the cospan
\begin{align}
\xymatrix@C=12ex
{
(H',\xi')~\ar@{>->}[r]^{m'} & CX \ar@<-1ex>[r]_{\left\langle\hat{c}_2\circ \hat{k}\right\rangle_{k\in K}} \ar@<3pt>[r]^{\left\langle \hat{c}_1\circ \hat{k}\right\rangle_{k\in K}} & \prod\limits_{k\in K} CS
}\label{diag:invariant}
\end{align}
where $K=\{k: UCX\to X\}$ is the set of $X$-colouring maps.  
Since $C$ is right-adjoint, it preserves products and $\prod_k CS\simeq C\prod_k S$. 
Thus, we are considering as corelation a pair of maps which can perform two `$S$-classifications' of an $X$-pattern \emph{and all its $X$-recolourings}, simultaneously.  
Clearly, $m'\ \hspace{2pt}\top\hspace{2pt} (H',\xi')$, so $(H',\xi')$ belongs to the covariety it defines.

By taking the cokernel pair of the equalizer $m'$ above $(H',\xi')\mono CX\rightrightarrows \coker(m')$ we get a corelation which is invariant by construction. In fact, we get an \emph{invariant cocongruence}, which are to cofree coalgebras what stable congruences are to free algebras.  We can now state the dual of \cref{thm:stablecongruence}:

\begin{theorem}
Let $T$ be a covarietor, let $c_1,c_2: UCX\rightrightarrows S$ be an invariant $T$-coequational specification,  and consider the equalizer
\[
\xymatrix
{
(H,\xi)~\ar@{>->}[r]^{m} & CX\ar@<-1pt>[r]_{\hat{c}_1} \ar@<3pt>[r]^{\hat{c}_2} & CS.
}
\]
Then $m\coperp (H,\xi)$. Conversely, if $m\coperp (H,\xi)$, then $\coker(m)$ is invariant.
\end{theorem}

\begin{example}
Let us consider the same endofunctor as in \cref{ex:monoids}, namely the functor $\Sigma X= X\times X+1$. 
This functor is a covarietor and the cofree $\Sigma$-coalgebra $C_\Sigma X$ over $X$ is the set of finite and infinite binary trees whose nodes are labelled by the elements of $X$ \cite{adamek2003varieties}.  Consider in particular the cofree $\Sigma$-coalgebra over a set of two colours, which we shall write as $\{b,w\}$ for `black' and `white'.  We now define the coequation $c_1,c_2: UC\{b,w\}\rightrightarrows 2$
\begin{mathpar}
c_1(t)=\begin{cases}
1 & \text{if }\mathsf{LeftChild}(t)\text{ is labelled }b \\
0 & \text{else}
\end{cases}
\and
c_2(t)=\begin{cases}
1 & \text{if }\mathsf{RightChild}(t)\text{ is labelled }b \\
0 & \text{else}
\end{cases}
\end{mathpar}
where $\mathsf{LeftChild}(t)$ being labelled $b$ assumes that it exists, \ie that $t$ is not a leaf state, and similarly for  $\mathsf{RightChild}(t)$.  
Recall that a \emph{coequation-as-corelation} defines the set of behaviours which cannot be distinguished by the two classification schemes. 
The coequation above defines the covariety of finite and infinite binary trees with the property that if a state has left and a right children states, then they must be equal; \ie the covariety of deterministic binary trees. 
To see this, consider the equalizer $m:(H,\xi)\to CX$ of $\hat{c}_1,\hat{c}_2$.
It contains all binary trees such that left- and right-successors share a colour. 
Its cokernel defines the cocongruence on $CX+CX\to C\{b,w\}$, which only identifies $(1,s)$ and $(2,t)$ if $s=t$ belongs to $H$. 
This cocongruence is not invariant: the two copies in $CX+CX$ of the left-hand tree $t_l$ below are identified by the corelation $c_1,c_2$ (and its coreflexive closure), but can be recoloured into two copies of the right-hand tree $t_r$, which will be kept distinct by the corelation.
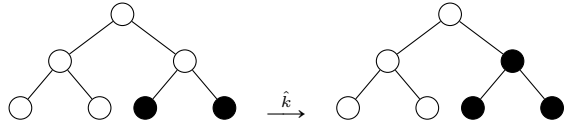
\begin{center}
\footnotesize{
\begin{tikzpicture}
\tikzset{
  node/.style = {align=center, inner sep=0pt}
}
\node [circle,draw] (1){};
\node [circle,draw, below left=4mm and 6mm of 1] (2){};
\node [circle,draw, below right=4mm and 6mm of 1] (3){};
\node [circle,draw,below left=4mm and 3mm of 2] (4){};
\node [circle,draw,below right=4mm and 3mm of 2] (5){};
\node [circle,draw,below left=4mm and 3mm of 3,fill=black] (6){};
\node [circle,draw,below right=4mm and 3mm of 3,fill=black] (7){};
\path (1) edge (2);
\path (1) edge (3);
\path (2) edge (4);
\path (2) edge (5);
\path (3) edge (6);
\path (3) edge (7);
\end{tikzpicture}
\quad
$\stackrel{\hat{k}}{\longrightarrow}$
\quad
\begin{tikzpicture}
\tikzset{
  node/.style = {align=center, inner sep=0pt}
}
\node [circle,draw] (1){};
\node [circle,draw, below left=4mm and 6mm of 1] (2){};
\node [circle,draw, below right=4mm and 6mm of 1,fill=black] (3){};
\node [circle,draw,below left=4mm and 3mm of 2] (4){};
\node [circle,draw,below right=4mm and 3mm of 2] (5){};
\node [circle,draw,below left=4mm and 3mm of 3,fill=black] (6){};
\node [circle,draw,below right=4mm and 3mm of 3,fill=black] (7){};
\path (1) edge (2);
\path (1) edge (3);
\path (2) edge (4);
\path (2) edge (5);
\path (3) edge (6);
\path (3) edge (7);
\end{tikzpicture}}
\end{center}
By constructing the invariant closure of the corelation using the construction of \eqref{diag:invariant}, the two trees above become components in the tuple of all recolourings of $t_l$. Applying $\hat{c}_1$ and $\hat{c}_2$ component-wise to this tuple will yield two tuples which will disagree at the coordinate of $t_r$. In fact, the equalizer $(H',\xi')$ of $\langle\hat{c}_1\circ \hat{k}\rangle_{k\in K},\langle\hat{c}_2\circ \hat{k}\rangle_{k\in K}: C\{b,w\}\rightrightarrows\prod_k C2$ contains precisely the trees whose nodes at depth $n$ all have the same colour, in other words the equalizer is isomorphic to $\{b,w\}^*\cup\{b,w\}^\omega$. From this it follows that given an arbitrary $\Sigma$-coalgebra $(V,\gamma)$,  if \emph{any} recolouring map $\hat{k}: (V,\gamma) \to C\{b,w\}$ has to factor through $\{b,w\}^*\cup\{b,w\}^\omega$
it must be the case that $\pi_1(\gamma(x))=\pi_2(\gamma(x))$ or $\gamma(x)\in 1$ for all $x\in V$.

\end{example}


\section{Coequations-as-predicates}\label{sec:predicates}

The coequations-as-predicates paradigm provides a picture of coequations that is very flexible with regard to how they can be written.
In this paradigm, a coequation is a subset of a cofree coalgebra, so any method of describing subsets can be used.
In practice, the elements of a cofree coalgebra carry some structure, for eg. a tree or a stream of numbers, allowing the user to describe coequations in terms of this structure.

In \cref{sub:behavioural coequations} and \cref{sub:beyond behaviour}, we will see some examples of predicate coequations and their descriptions.
We then give a brief account of Ad\'amek and Schwencke's observation that there is an inherent logical structure to coequations in \cref{sub:logic and avoiding patterns}.
Finally, in \cref{sub:beyond sets}, we talk about the expressiveness of predicate coequations in general, and give a generalization of predicate coequations when there are no cofree coalgebras.

In this section, we entirely focus on coalgebras in \(\Set\).
Many of the results can be generalized to coalgebras over other base categories, see for example \cite{adamek2003varieties} and the thesis \cite{hughes2001study}.

\subsection{Behavioural coequations}\label{sub:behavioural coequations}


Fix a covarietor \(T\) on \(\Set\) with forgetful-cofree adjunction \(U \dashv C\), and let \((Z, \delta)\) denote the final coalgebra \(C1\).
Given two coalgebras $(V_1, \gamma_1), (V_2,\gamma_2)$, two states \(v_1 \in V_1\) and \(v_2 \in V_2\) are said to be \emph{behaviourally equivalent} (\eg  \cite{kupkeleal2009behaviour}) if there is a third coalgebra \((V', \gamma')\) and homomorphisms \(h_i : V_i \to V'\) such that \(h_1(v_1) = h_2(v_2)\).
Behavioural equivalence is an equivalence relation: it is reflexive and symmetric, and since the category \(\Coalg\) has pushouts, it is also transitive.
Furthermore, since every coalgebra \((V, \gamma)\) admits a unique coalgebra homomorphism \(!_V : (V, \gamma) \to (Z, \delta)\), the state \(!_V(v)\) of \(Z\) is a representative of the behavioural equivalence class of \(v\) for any state \(v \in V\).
This is the motivation for calling the states of \((Z, \delta)\) \emph{behaviours} (for the functor $T$).

A \emph{behavioural coequation} is a subset of \(Z\).
A coalgebra \((V, \gamma)\) \emph{satisfies} \(W \subseteq Z\), written \((V, \gamma) \models W\), if \(\im(!_V) \subseteq W\), \ie a coalgebra satisfies a behavioural coequation if the coequation contains all of the behaviours exhibited by the coalgebra.
A \emph{behavioural covariety} is a class of coalgebras of the form \(\Cov(W) = \{(V,\gamma) \mid (V,\gamma) \models W\}\) for some \(W \subseteq Z\). 

\begin{example}\label{eg:deterministic automata}
	Coalgebras for the functor \(T_{det} = 2 \times \Id^A\) are deterministic automata.
	The final deterministic automaton is \((2^{A^*}, \langle\epsilon?,\partial\rangle)\), where \(A^*\) is the set of empty  or nonempty words in the alphabet \(A\) (here, \(\epsilon\) is the empty word),  and
	\[{
			\epsilon?(L) = \begin{cases}
				1 &\text{if}\ \epsilon \in L\\
				0 &\text{otherwise}
			\end{cases} \qquad 
			\partial(L)(a) = \{w \in A^* \mid aw \in L\}
		}\]
	for any \(L \subseteq A^*\) and \(a \in A\)~\cite{rutten1996universal,brzozowski1964derivatives}.
	Recall that the set \(\textsf{Reg} \subseteq 2^{A^*}\) of \emph{regular languages} is the smallest subset of \(2^{A^*}\) closed under concatenation, iteration, and finite unions, and containing \(\{\epsilon\}\) and \(\{a\}\) for each \(a \in A\).
	The class \(\Cov(\textsf{Reg})\) of deterministic automata that accept regular languages is a behavioural covariety.

	By Kleene's theorem, a deterministic automaton satisfies \(\textsf{Reg}\) if and only if it is bisimilar to a locally finite automaton.
	There are many examples of deterministic automata that satisfy \(\textsf{Reg}\) but are not locally finite: The following automata are bisimilar and both accept the language \(a^*\), but only one of them is locally finite.
	
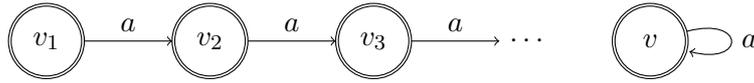
\[{\begin{tikzpicture}
			\node[state,accepting] (0) {\(v_1\)};
			\node[state,accepting, right=3em of 0] (1) {\(v_2\)};
			\node[state,accepting, right=3em of 1] (2) {\(v_3\)};
			\node[right=3em of 2] (3) {\(\cdots\)};
	
			\path (0) edge[above, ->] node{\(a\)} (1);
			\path (1) edge[above, ->] node{\(a\)} (2);
			\path (2) edge[above, ->] node{\(a\)} (3);
		\end{tikzpicture} 
		\qquad
		\begin{tikzpicture} 
			\node[state,accepting] (0) {\(v\)};
			\draw (0) edge[loop right, right, ->] node{\(a\)} (0); 
		\end{tikzpicture}}\]

\end{example}

The relationship between regular languages and finite automata in \cref{eg:deterministic automata} is typical of behavioural coequations.
Behavioural coequations constrain dynamics by constraining behaviour, and behaviour is preserved under many of the useful operations on coalgebras.
In general, behavioural coequations carve nicely structured categories out of \(\Coalg\).  Following the co-Birkhoff result of \cref{thm:coBirkhoff}, we will say that a class of coalgebras is a \emph{structural covariety} if it is closed under homomorphic images, subcoalgebras and coproducts.  Note that this concept makes sense whether $T$ is a covarietor or not.

\begin{proposition}[Rutten~\cite{rutten1996universal}]
	For any \(W \subseteq Z\),  \(
		\Cov(W)
	\)
	is a \emph{structural covariety}.
\end{proposition}

However, not every structural covariety is carved out of \(\Coalg\) by a behavioural coequation.
As we saw in \cref{eg:deterministic automata}, locally finite deterministic automata are not behaviourally specified: if they were, their defining coequation would be \(\textsf{Reg}\).
On the other hand, we will see in \cref{sub:beyond behaviour} that (under mild conditions) locally finite coalgebras form a structural covariety.

The mismatch between behavioural coequations and covarieties does not detract from the importance of behavioural constraints in the computer science literature.
Behavioural coequations are particularly common in fields like automata theory and process algebra where specification languages play an important role.

\begin{example}\label{eg:process algebra example}
	Fix a set \(A\) of \emph{atomic actions}, and consider the following BNF grammar
	\begin{mathpar}
		E ::= a \in A \mid x \in \text{Var} \mid E + E \mid a(E) \mid \mu x.F \and F ::= a \in A  \mid F + F \mid a(E) \mid \mu x.F
	\end{mathpar}
	A \emph{specification} is an expression \(e \in E\) in which every variable \(x \in \text{Var}\) appears within the scope of a \(\mu x\). 
	The set of specifications can then be given the structure of a \(\mathcal P_{\omega}(\{\checkmark\} + \Id)^A\)-coalgebra \((\text{Exp}, \partial)\) using the GSOS law below.
	For any \(a \in A\), \(e,e_1,e_2,f \in \text{Exp}\), infer
	\begin{mathpar}
		\infer{a \trans{a} \checkmark}{\ \ }
		\and
		\infer{a(e) \trans{a} e}{\ \ }
		\and
		\infer{e_i \trans{a} f}{e_1 + e_2 \trans{a} f}
		\and
		\infer{f[\mu x. f/x] \trans{a} e}{\mu x.f \trans{a} e}
	\end{mathpar}
	Here, \(e \trans{a} \xi\) means that \(\xi \in \partial(e)(a)\).
	The functor \(\mathcal P_\omega(\{\checkmark\} + \Id)^A\) has a final coalgebra \((Z, \delta)\) consisting of \(A\)-decorated trees with transitions carrying labels from \(A\) and leaves carrying the label \(\checkmark\)~\cite{worrell1999terminal}.
	Every specification \(e\) gives rise to a unique behaviour \(!_{\text{Exp}}(e)\), and therefore also a tree.
	In analogy with regular languages (see \cref{eg:deterministic automata}), one might call the set of behaviours arising from specifications the \emph{regular} coequation.
	A process satisfies the regular coequation if every of its states mimics the behaviour of a specification.

	Many pairs of specifications give rise to identical behaviours: For example, \(e + f\) and \(f + e\) are behaviourally equivalent for any \(e\) and \(f\), and so are \(f[\mu x. f/x]\) and \(\mu x.f\). 
	Studying behavioural equivalences like these is a popular topic in process algebra~\cite{fokkink2013introduction}.
\end{example}

The reader familliar with process algebra should note that in many cases, including \cref{eg:process algebra example}, behavioural equivalence and \emph{bisimilarity} coincide.
For a general \(T\), a bisimulation between \(T\)-coalgebras \((V_1, \gamma_1)\) and \((V_2, \gamma_2)\) consists of a coalgebra \((R, \rho)\) and a pair of coalgebra homomorphisms \(\pi_i : (R, \rho) \to (V_i, \gamma_i)\). 
Image factorisations exist in \(\Set\), so every bisimulation is equivalent to one in which \(R \subseteq V_1 \times V_2\) and the homomorphisms \(\pi_1,\pi_2\) are the projections of \(R\) onto the first and second components of \(R\)~\cite{rutten1996universal}.
If two states \(v_1 \in V_1,v_2 \in V_2\) are related by a bisimulation, we say that \(v_1\) and \(v_2\) are \emph{bisimilar} and write \(v_1 \bisim v_2\).
Important examples of bisimulations include graphs of homomorphisms: 
in fact, a function \(V_1 \to V_2\) is a coalgebra homomorphism if and only if its graph is a bisimulation~\cite{rutten1996universal,gumm2001products}.

\begin{lemma}[Rutten~\cite{rutten1996universal}]
	Let \((V_1, \gamma_1)\) and \((V_2, \gamma_2)\) be a coalgebras, and \(v_1 \in V_1\) and \(v_2 \in V_2\) be states.
	If \(T\) preserves \emph{weak pullbacks},
	 then \(v_1 \bisim v_2\) if and only if \(v_1\) and \(v_2\) are behaviourally equivalent. 
\end{lemma}

Many of the covarietors familliar to computer scientists preserve weak pullbacks, including every polynomial functor, the covariant powerset functor \(\mathcal P\) and its $\kappa$-accessible variants \(\mathcal P_\kappa\), and every product, coproduct, and composition of these functors~\cite{gumm1999elements}.
Among the resulting class of functors are the deterministic automaton functor \(B \times \Id^A\) and the nondeterministic automaton functor \(B \times \mathcal P(\Id)^A\) for any output set \(B\).
The added assumption that \(T\) preserve weak pullbacks leads to a rich coalgebraic theory, and much of \cite{rutten1996universal} depends on it.

Under the additional assumption that \(T\) preserves weak pullbacks, Gumm and Schr\"o\-der obtain the following characterisation of behavioural covarieties.

\begin{theorem}[Gumm and Schr\"oder~\cite{gumm1998covarieties}]\label{thm:total bisimilarity theorem}
	Let \(T\) preserve weak pullbacks.
	Then a structural covariety \(\catfont C\) in \(\Coalg\) is behavioural if and only if it is \emph{closed under total bisimulations}, \ie if \((V_1, \gamma_1) \in \catfont C\) and there is a bisimulation \((R, \rho)\) between \((V_1, \gamma_1)\) and \((V_2,\gamma_2)\) such that \(\pi_1\) and \(\pi_2\) are surjective, then \((V_2, \gamma_2) \in \catfont C\) as well. 
\end{theorem}

\begin{example}\label{eg:alt}
	Consider the functor \(T_{alt} = (\{\checkmark\} + \Id)^{\{a,b\}}\).
	In a \(T_{alt}\)-coalgebra \((V, \gamma)\), write \(v\trans{a} v'\) if \(v' = \gamma(v)(a)\) and \(v \Rightarrow a\) if \(\checkmark = \gamma(v)(a)\), and similarly for \(b\).
	The functor \(T_{alt}\) preserves weak pullbacks,  and the final \(T_{alt}\)-coalgebra is the set of all \(\{a,b\}\)-decorated trees \(t\) such that every node \(n\) of \(t\) has at most one \(n \trans{a} n'\) transition and at most one \(n \trans{b} n'\), and all other transitions are of the form \(n \Rightarrow a\) or \(n \Rightarrow b\).
	The coalgebra structure \(\delta\) is the obvious parent-child transition structure.

	The class of \emph{sequence} coalgebras, consisting all those \(T_{alt}\)-coalgebras \((V, \gamma)\) such that \(|\gamma^{-1}(\checkmark)| = 1\), is closed under total bisimulations.
	By \cref{thm:total bisimilarity theorem}, the covariety of sequence coalgebras is determined by a set \(W_{seq}\) of behaviours.
	One way to describe this coequation is as follows: \(W_{seq}\) is the set of all behaviours \(t\) such that the first \emph{layer} of \(t\) is one of 
	
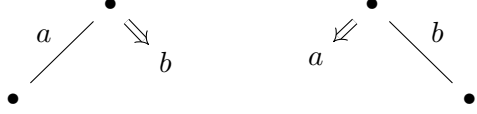
\[\begin{tikzpicture}
		\node (0) {\(\bullet\)};
		\node[below left=3em of 0] (1) {\(\bullet\)};
		\node[below right=1em of 0] (2) {\(b\)};
		\path (0) edge[above left] node{\(a\)} (1);
		\path (0) edge[double, double distance=2pt, -implies] (2);
	\end{tikzpicture}
	\qquad\qquad
	\begin{tikzpicture}
		\node (0) {\(\bullet\)};
		\node[below left=1em of 0] (1) {\(a\)};
		\node[below right=3em of 0] (2) {\(\bullet\)};
		\path (0) edge[above right] node{\(b\)} (2);
		\path (0) edge[double, double distance=2pt, -implies] (1);
	\end{tikzpicture}\]\vspace{-2em}\\
	Of course, many of the trees in \(W_{seq}\) are \emph{not} behaviours exhibited by sequence \(T_{alt}\)-coalgebras, as nodes from deeper layers might accept too many or too few of \(a,b\).
	This can easily be fixed once we have made the following observation.
\end{example}

\begin{lemma}[Rutten~\cite{rutten1996universal}]
 	Let \((V, \gamma),(V',\gamma')\) be \(T\)-coalgebras and \(h : V \to V'\) be a \(T\)-coalgebra homomorphism.
	Then \(h(V)\) is a subcoalgebra of \((V', \gamma')\).
\end{lemma} 

Consequently, if \((V, \gamma) \models W\) for some \(W \subseteq Z\), then \(!_V(V)\) is a subcoalgebra of \((Z, \delta)\) contained in \(W\).
It follows from \(\Coalg\) being cocomplete and having image factorisations that an arbitrary union of subcoalgebras of a fixed coalgebra is a subcoalgebra.
In particular, there is a largest subcoalgebra \(\Box W\) contained in \(W\), and it satisfies
\(
	\Cov(\Box W) = \Cov(W)
\)
for any \(W \subseteq Z\).
In fact, \(\Box W\) is the final object of \(\Cov(\Box W)\)!
The operator \(\Box\) is known as the ``henceforth'' operator in \cite{jacobs2002temporal}, and is studied in more general settings by Hughes in \cite{hughes2001modal}.
	
\subsection{Beyond behaviour}\label{sub:beyond behaviour}



A \emph{predicate coequation in \(X\) colours} is a set of \emph{\(X\)-patterns}, or a subset of the cofree coalgebra \(CX\).
A coalgebra \((V, \gamma)\) \emph{satisfies} the predicate coequation \(W \subseteq CX\) if, for any colouring \(c : V \to X\), the homomorphism \(\hat c : (V, \gamma) \to CX\) induced by the adjunction \(U \dashv C\) factors through \(W\) (that is, \(\hat c(V) \subseteq W\)). 

\begin{theorem}[Rutten~\cite{rutten1996universal}, Gumm~\cite{gumm1999elements}]\label{thm:closure for predicates 1}
	Let \(T\) be a covarietor, \(X\) be a set, and \(W \subseteq UCX\).
	The class \(
			\Cov(W) = \{(V,\gamma) \mid (\forall c : V \to X)\ \hat c(V) \subseteq W\}
		\) 
	is closed under subcoalgebras, coproducts, and homomorphic images.
\end{theorem}

Under mild assumptions on \(T\), every structural covariety is presentable by a coequation.
The key to proving the converse to \cref{thm:closure for predicates 1} is to give an upper bound on the number of colours that only depends on \(T\).
This is possible when \(T\) is \emph{\(\kappa\)-bounded} for some cardinal \(\kappa\), meaning that for any coalgebra \((V, \gamma)\) and any state \(v \in V\), there is a subcoalgebra \(S\) of \((V,\gamma)\) such that \(v \in S\) and \(|S| \le \kappa\).
An endofunctor on \(\Set\) is \emph{bounded} if it is \(\kappa\)-bounded for some \(\kappa\).\footnote{Equivalently, \(T\) is \emph{accessible} or \emph{small}~\cite{adamek2001varieties}.}
The class of bounded functors is broad enough to capture most functors in everyday use by computer scientists~\cite{rutten1996universal,gumm2001functors}.
At the beginning we assumed that \(T\) is a covarietor, but boundedness actually implies this. 

\begin{theorem}[Kawahara \& Mori~\cite{kawaharamori2000small}]
	If \(T : \Set \to \Set\) is bounded, then \(T\) is a covarietor.
\end{theorem}

Let \(\catfont C\) be a structural covariety and \(\kappa\) an infinite cardinal, and assume that \(T\) is \(\kappa\)-bounded.
Given a state \(v\) of a coalgebra \((V,\gamma)\), there is a subcoalgebra \(S\) of \((V,\gamma)\) containing \(v\) such that \(|S| \le \kappa\).
By a simple renaming of states, \(S\) is isomorphic to a coalgebra whose state space is a set of numbers in \(\kappa\).
This means that every \(T\)-coalgebra is locally isomorphic to a coalgebra of the form \((S, \sigma)\) where \(S \subseteq \kappa\), and in particular that every coalgebra in \(\catfont C\) is locally of this form.
Writing \(\catfont G = \{(S,\sigma) \in\catfont C \mid S \subseteq \kappa\}\), the coequation 
\(
	W_{\catfont G} = \bigcup \{\hat c(S) \mid \text{\((S,\sigma) \in \catfont G\) and \(c : S \to \kappa\)}\}
\) 
determines \(\catfont C\), meaning that \(\catfont C = \Cov(W_{\catfont G})\).

\begin{theorem}[Rutten~\cite{rutten1996universal}]\label{thm:rutten cobirkoff}
	If \(T\) is \(\kappa\)-bounded and \(\catfont C\) is a structural covariety in \(\Coalg\), then \(\catfont C = \Cov(W)\) for some \(W \subseteq C\kappa\).
\end{theorem}

Behavioural coequations are instances of predicate coequations: They are the coequations in \(1\) colour.
We have already seen in \cref{eg:deterministic automata,eg:process algebra example,eg:alt} that some covarieties are presentable by behavioural coequations despite the type functor failing to be \(1\)-bounded.\footnote{It is straightforward to check that each is \(\omega\)-bounded, on the other hand.}
The bound on the number of colours provided by \cref{thm:rutten cobirkoff} is rarely optimal in practice.

Towards a better bound, recall the definition of behavioural equivalence from \cref{sub:behavioural coequations}.
Let \(\catfont C\) be a covariety that is \emph{closed under behavioural equivalence}, meaning that it contains every coalgebra \((V,\gamma)\) such that every state \(v \in V\) is behaviourally equivalent to a state of a coalgebra in \(\catfont C\).
Then \(\catfont C\) is necessarily behavioural.
Indeed, if \(W_{\catfont C} = \{!_{S}(x) \mid x \in S, (S,\sigma) \in \catfont C\}\), then \((S,\sigma) \models W_{\catfont C}\) for any \((S,\sigma) \in \catfont C\).
Conversely, every state of a coalgebra satisfying \(W_{\catfont C}\) behaves like a state from a coalgebra in \(\catfont C\), and by assumption such a coalgebra must be in \(\catfont C\).

This argument generalizes to the \(\lambda\)-pattern situation by saying that a class \(\catfont C\) of coalgebras is \emph{closed under \(\lambda\)-pattern equivalence} if \((V,\gamma) \in \catfont C\) whenever the following condition is met: for any \(v \in V\) and any colouring \(c : V \to \lambda\), there is a \((S,\sigma) \in \catfont C\) with a state \(x \in S\) and a colouring \(c' : S \to \lambda\) such that \(\hat c(v) = \hat c'(x)\). We thus have:

\begin{proposition}\label{prop:first actual bound}
	Let \(\catfont C\) be a covariety closed under \(\lambda\)-pattern equivalence.
	Then \(\catfont C = \Cov(W)\) for some \(W \subseteq UC\lambda\).
\end{proposition}

\begin{example}
	Consider the deterministic automaton endofunctor \(T_{det} = 2 \times \Id^A\) from \cref{eg:deterministic automata}.
	Fix two words \(w_1,w_2 \in A^*\) and let \(\catfont C\) be the class of deterministic automata \((V,\gamma)\) in which \(v \trans{w_1} v'\) and \(v \trans{w_2} v''\) implies \(v' = v''\).
	Consider an automaton \((V,\langle o,\partial\rangle)\), \(v \in V\), and \(c : V \to 2\) the colouring \(c(x) = 1 \iff x = \partial(v)(w_1)\),\footnote{Here, \(\partial(v) : A^* \to V\) is defined by \(\partial(v)(\epsilon) = v\) and \(\partial(v)(wa) = \partial(\partial(v)(w))(a)\).}  and assume there is an automaton \((V',\langle o',\partial'\rangle) \in \catfont C\) with \(v' \in V'\) and a colouring \(c' : V' \to 2\) such that \(\hat c(v) = \hat c'(v')\).
	Then \(c(\partial(v)(w_2)) = c'(\partial'(v')(w_2)) = c'(\partial'(v')(w_1)) = c(\partial(v)(w_1)) \), so that \(\partial(v)(w_2) = \partial(v)(w_1)\) by construction of \(c\).
	It follows that \(\catfont C\) is closed under \(2\)-pattern equivalence, and is therefore determined by a coequation in \(2\) colours.
	Indeed, where \(\varepsilon\) is the counit, it is given by
	\[
	W = \{t \in UC2 \mid \varepsilon_2(\partial(t)(w_1)) = \varepsilon_2(\partial(t)(w_2))\}.
	\] 
\end{example}

In \cref{sub:behavioural coequations}, we saw that bisimilarity and behavioural equivalence coincide when \(T\) preserves weak pullbacks.
This gave way to \cref{thm:total bisimilarity theorem}, which characterised behavioural covarieties in terms of total bisimilarity.
Ad\'amek applies the same reasoning in \cite{adamek2005logic} to give a bound like the one in \cref{prop:first actual bound} in terms of bisimulations:
For a cardinal \(\lambda\), say that a covariety \(\catfont C\) is \emph{closed under \(\lambda\)-colour bisimilarity} if \((V, \gamma) \in \catfont C\) whenever the following condition is met: For any \(c : V \to \lambda\) there is a \((S, \sigma) \in \catfont C\), a colouring \(c' : V' \to \lambda\), and a total bisimulation \((R, \rho)\) between \((V, \gamma)\) and \((V', \gamma')\) such that \(c \circ \pi_1 = c' \circ \pi_2\).

\begin{theorem}[Ad\'amek~\cite{adamek2005logic}]\label{thm:Adamek n-colour}
	Suppose \(T\) is a covarietor that preserves weak pullbacks.
	A covariety in \(\Coalg\) is presentable by a predicate coequation in \(\lambda\) colours if and only if it is closed under \(\lambda\)-colour bisimilarity.
\end{theorem}


\begin{example}\label{eg:simple graphs}
	The class \(\catfont C\) of \emph{simple} graphs (of finite degree) can be seen as a covariety in \(\Coalg[\pow_{\omega}]\).
	In a \(\pow_\omega\)-coalgebra \((V, \gamma)\), write \(v_1 \to v_2\) to denote \(v_2 \in \gamma(v_1)\).
	Then \((V, \gamma)\) is a \emph{simple graph} if \(\to\) is a reflexive symmetric relation on \(V\).\footnote{A simple graph in this sense contains the same information as the more traditional concept from combinatorics. However, \(\mathcal P_{\omega}\)-coalgebra homomorphisms are not graph homomorphisms. For a coalgebraic depiction of traditional directed graphs, see pg. 22 of \cite{rutten2000universal}, or \cite{jkel2015unified}.}
	Reflexivity and symmetry are given by the coequations
	\(W_{ref} = \{t \mid (\exists s)\ \varepsilon_2(s) = \varepsilon_2(t)\ \text{and}\ t \to s\}\)
	and
	\(W_{sym} = \{t \mid (\forall s)\ (t \to s \implies (\exists u)\ \varepsilon_2(u) = \varepsilon_2(t)\ \text{and}\ s \to u)\}\)
	respectively. 
	The modal logician will recognise the axioms (T) $p\to\diamondsuit p$ and (B) $p\to\Box\diamondsuit p$, the roles of the propositional variable $p$ being played by the colouring map $\varepsilon_2$. 
	The coequation we are looking for is therefore \(W_{sim} = W_{ref}\cap W_{sym}\).
	By \cref{thm:Adamek n-colour}, \(\catfont C\) is closed under \(2\)-colour bisimilarity.
	This covariety is also not behavioural: \(\bullet \righttoleftarrow\) and \(\bullet \to \bullet \to \bullet \to \cdots\) are bisimilar, for example.
\end{example}

\begin{example}
	Recall that a coalgebra \((V, \gamma)\) is \emph{locally finite} if for any \(v \in V\), there is a subcoalgebra \(S\) of \((V,\gamma)\) such that \(v \in S\) and \(S\) is finite.
	If \(T\) preserves weak pullbacks, then the class \(\catfont C_{\omega}\) of locally finite \(T\)-coalgebras is a covariety in \(\omega\) colours.

	To see why, let \((V,\gamma)\) be a coalgebra such that for any \(c : V \to \omega\), there is a locally finite \((V',\gamma')\), a colouring \(c' : V' \to \omega\), and a total bisimulation \((R,\sigma)\) between \((V,\gamma)\) and \((V',\gamma')\) such that \(c \circ \pi_1 = c' \circ \pi_2\).
	If \(v \in V\) and \(c : V \to \omega\) is any colouring, and we take \((V',\gamma'),(R,\rho),c:V' \to \omega\) as before, then \(v R v'\) for some \(v' \in V'\).
	Since \((V',\gamma')\) is locally finite, \(c'(S')\) is finite for some subcoalgebra \(S'\) of \((V',\gamma')\) containing \(v'\). 
	The projection \(\pi_2\) is surjective, so \(P = \pi_2^{-1}(S')\) is a subcoalgebra of \((R,\rho)\) containing \((v,v')\).
	Taking images, we see that \(
			c(\pi_1(P)) = c'(\pi_2(P)) = c'(S'),
		\)
	so that \(\pi_1(P)\) is a subcoalgebra of \((V,\gamma)\) with finite image under \(c\).
	Since \(T\) preserves weak pullbacks, there is a smallest subcoalgebra \(\langle v\rangle\) of \((V,\gamma)\) containing \(v\)~\cite{rutten1996universal}.
	This subcoalgebra is contained in every \(P\) as constructed above, so has finite image under \(c\) for any \(c\).
	A set is finite if and only if every image of the set under a map into \(\omega\) has a finite image, so \(\langle v\rangle\) must be finite.
	It follows that \((V,\gamma)\) is locally finite, so by \cref{thm:Adamek n-colour}, \(\catfont C_{\omega}\) is presentable with a coequation in \(\omega\) colours.
	The desired coequation consists of those \(\omega\)-patterns in which only finitely many colours appear.
\end{example}

\subsection{Logic and Avoiding Patterns}\label{sub:logic and avoiding patterns}

As we have already seen, it is possible for distinct coequations, like \(\Box W\) and \(W\) in \cref{eg:alt} for instance, to specify the same covariety.
In this short section, we describe Ad\'amek and Schwencke's framing of this equivalence between coequations as a logical equivalence in \cite{adamek2005logic,schwencke2008coequational,schwencke2010coequational}, and discuss Adam\'ek's sound and complete deduction system for the resulting logic of coequations for a polynomial endofunctor \(T\).

Given two coequations \(W_1\) and \(W_2\), \(W_1\) is said to \emph{imply} \(W_2\), written \(W_1 \models W_2\), if for any coalgebra \((V,\gamma)\), \((V,\gamma) \models W_2\) whenever \((V,\gamma) \models W_1\).
For example, \(W_1 \subseteq W_2\) implies \(W_1 \models W_2\), and \(W \models\Box W\) and \(\Box W \models W\).
The inference relation \(\models\) also interacts with recolourings: every \(h : X \to X\) induces \(\hat c : CX \to CX\) such that \(W \models \hat h(W)\).

Further analysis of the inference relation \(\models\) is possible with a notation used by Gumm.
In \cite{gumm1999elements}, Gumm gives a negative description of coequations, as predicates of the form \(\boxminus t = (CX) - \{t\}\) for a pattern \(t\). 
This is a particularly useful notation for coequations when patterns are easily described but general predicates are not.
Such is the case when \(T\) is a polynomial functor, as patterns are identifiable with certain trees.

Fix a polynomial functor \(T_\Sigma = \bigcup_{p \in \Sigma} \Id^{ar(p)}\), where \(\kappa\) is a cardinal, \(\Sigma\) is a set, and \({ar} : \Sigma \to \kappa\).
For a set of colours \(X\), an \(X\)-pattern is a tree \(t\) in which every node \(n\) is labelled with a pair \((p, x) \in \Sigma \times X\) and a transition function that maps each \(\alpha < ar(p)\) to each child of \(n\). 
The structure map of \(CX\) is given by parent\(\to\)child transitions:
if \(n\) is a node of \(t\) and \(n'\) is the \(\alpha\)th child of \(n\), then \(t \trans{\alpha} s\) when \(s\) is the subtree of \(t\) rooted at \(n'\).
There is a unique node of \(t\) with no incoming transitions, called its \emph{root}.
Each node of \(t\) is the root a tree, and we call trees of this form \emph{subtrees} of \(t\).
We write \(s \sqsubseteq t\) to denote that \(s\) is a subtree of \(t\). 

Given \(t,s \in UCX\), if \(s \sqsubseteq t\) and \((V,\gamma) \models \boxminus s\), then \((V,\gamma) \models \boxminus t\) as well.
This is because, if \(\hat c(v) = t\) for some colouring \(c : V \to X\) and \(v \in V\), then every path \(t \trans{\alpha_1} t_1 \trans{\ } \cdots \trans{\ } t_{n-1} \trans{\alpha_n} s\) is witnessed in \((V,\gamma)\) by a path \(v \trans{\alpha_1} v_1 \trans{\ } \cdots \trans{\ } v_{n-1} \trans{\alpha_n} u\) such that \(\hat c(u) = s\).\footnote{Here, \(v \trans{\alpha} u\) in \((V,\gamma)\) if \(u = \gamma(v)(\alpha)\).}
Furthermore, if \(s\) is a \emph{recolouring} of \(t\), \ie \(s = \hat k(t)\) for some colouring \(k : UCX \to X\), then \((V,\gamma) \models \boxminus s\) implies \((V,\gamma) \models \boxminus t\) as well.
This is due to the composition \(c' = k \circ \hat c\), where \(c : V \to X\) is any colouring of \((V,\gamma)\), since \(s = \hat c'(v)\) when \(t = \hat c(v)\).
We obtain the following proof rules.
\[
	\infer[\textbf{child}]{\vdash \boxminus t}{\vdash \boxminus s \qquad t \to s}
	\qquad
	\infer[\textbf{\(k\)-rec}]{\vdash \boxminus t}{\vdash \boxminus s \qquad s = \hat {k}(t)}
\]
For any \(t,s \in UCX\), if \(\vdash \boxminus t\) can be deduced from \(\vdash \boxminus s\) with \textbf{\(k\)-rec} (here, \(k\) is allowed to vary) and \textbf{child}, we write \(\boxminus s \vdash \boxminus t\).

\begin{theorem}[Ad\'amek~\cite{adamek2005logic}]
	For a polynomial endofunctor \(T_\Sigma\), a set \(X\), and any \(s,t \in CX\), \(\boxminus s \models \boxminus t\) if and only if \(\boxminus s \vdash \boxminus t\).
\end{theorem}

As shown in Ad\'amek's~\cite{adamek2005logic} and Schwencke's \cite{schwencke2008coequational,schwencke2010coequational}, the logic of coequations for polynomial functors (described above) can be extended to include many bounded functors.
The extended logic relies on the fact that every bounded functor is a natural quotient of some polynomial functor~\cite{adamekporst2004tree}, and by extension every cofree coalgebra for a bounded functor is a quotient of a cofree coalgebra for a polynomial functor.
The subtree and recolouring rules apply to representatives, and with the right natural quotient\footnote{Namely, a so-called \emph{regular presentation}. See \cite{schwencke2010coequational} for details.} \(T_\Sigma \Rightarrow T\), the ensuing logic is sound and complete with respect to coequational reasoning.  

\subsection{Generalized coequations}\label{sub:beyond sets}


If \(T\) is not bounded, \(T\) is likely not a covarietor.
In such a case, we cannot always use predicates to specify classes of coalgebras over the base category \(\Set\).
However, as Aczel and Mendler showed in \cite{aczel1989final}, every endofunctor on \(\Set\) extends to a covarietor on the category of classes.
By \emph{approximating} cofree coalgebras, which may be proper classes in the case that \(T\) is unbounded, Ad\'amek recovers \emph{generalized coequations} in \cite{adamek2005birkhoff}, and shows they are sufficient for specifying structural covarieties in general.

To approximate the cofree coalgebra in \(X\) colours, we follow Barr in \cite{barr1993terminal} and construct its \emph{final sequence}, the ordinal-indexed diagram
\[\xymatrix{
	X_0 & X_1\ar@{->}[l]_{\phi_0^1} & X_2\ar@{->}[l]_{\phi_1^2} & \cdots \ar@{->}[l] & X_\omega \ar@{->}[l] & X_{\omega + 1} \ar@{->}[l]_{\phi_\omega^{\omega+1}} & \cdots \ar@{->}[l]
}\]
Here, \(X_0 = 1\) and \(\phi_0^1 = {!}\), \(X_{\alpha + 1} = X \times TX_\alpha\), and \(\phi_{\alpha + 1}^{\beta+1} = \id_X \times T(\phi_{\alpha}^\beta)\) for any ordinals \(\alpha < \beta\), and
\(
	(X_{\lambda}, \{\phi_{\alpha}^{\lambda}\}_{\alpha < \lambda}) = \varprojlim \{\phi_{\alpha}^\beta : X_\beta \to X_\alpha \mid \alpha < \beta < \lambda\}
\)
for \(\lambda\) a limit ordinal.

For any \(T\)-coalgebra \((V, \gamma)\) and any colouring \(c : V \to X\), let \(c_0 = {!} : V \to X_0\) be the unique such function, and define \(c_{\alpha+1} = \langle c, T(c_\alpha) \circ \gamma\rangle\) at successor ordinals, and \(c_{\lambda} : V \to X_{\lambda}\) to be the unique cone homomorphism \(\{c_{\alpha}\}_{\alpha < \lambda} \to \{\phi_\alpha^\lambda\}_{\alpha < \lambda}\) when \(\lambda\) is a limit ordinal.
A \emph{generalized \(X\)-pattern} is an ordinal indexed sequence \(\{t_\alpha\}_{\alpha \in \catfont{Ord}}\) such that \(t_\alpha \in X_\alpha\), and \(t_\alpha = \phi_{\alpha}^\beta(t_{\beta})\) for any \(\alpha < \beta\). 
A \emph{generalized coequation} is a class \(W\) of generalized patterns, and \((V,\gamma) \models W\) if for any \(c : V \to X\) and \(v \in V\), we find \(\{c_\alpha(v)\}_{\alpha \in \catfont{Ord}} \in W\).

\begin{theorem}[Ad\'amek~\cite{adamek2005birkhoff}]\label{thm:adamek covariety theorem}
	For any endofunctor \(T\), a class \(\catfont C\) of \(T\)-coalgebras is a structural covariety if and only if there is a generalized coequation \(W\) such that \(\catfont C = \Cov(W)\).
\end{theorem}

Generalized coequations are indeed generalisations of coequations for a covarietor.
If \(T\) is a covarietor, then \(\phi_\lambda^{\lambda + 1}\) is an isomorphism for some ordinal \(\lambda\)~\cite{adamek2003varieties}.\footnote{Worrell shows in \cite{worrell1999terminal} that if \(T\) is \(\kappa^+\)-bounded, \(\phi_\lambda^{\lambda+1}\) is an isomorphism when \(\lambda = \kappa + \kappa\).}
As \(\phi_{\lambda}^{\lambda + 1}\) is an isomorphism, it has an inverse \(\langle k,\delta^X\rangle : X_{\lambda} \to X \times TX_{\lambda}\), and the cofree coalgebra \(CX\) is precisely \((X_\lambda, \delta^X)\).
In this setting, \(c_\lambda\) and \(\hat c\) coincide, and the satisfaction relation from \cref{sub:beyond behaviour} coincides with the satisfaction of generalized coequations.
Note, however, that while the set of colours is fixed in \cref{thm:rutten cobirkoff}, the colours appearing in \cref{thm:adamek covariety theorem} can vary.

\begin{example}
	The powerset functor \(\mathcal P\) is not bounded, as no \(\phi_{\lambda}^{\lambda + 1}\) can be a bijection.
	Nevertheless, the class of simple graphs from \cref{eg:simple graphs} forms a covariety.
	The presenting coequation is \(2\)-coloured and can be visualised as a subset of \(X_3\).
	Here, \(X_3 = 2 \times \mathcal P(2 \times \mathcal P(2 \times 2))\), so elements of \(X_3\) can be thought of as extensional trees with \(2\)-coloured nodes and height at most \(3\).
	The desired subset, call it \(W\), is obtained by restricting the coequation in \cref{eg:simple graphs} to such trees.
	The generalized coequation \(W_{sim}\) then consists of all ordinal-indexed sequences \(\{t_\alpha\}_{\alpha \in \catfont{Ord}} \) such that $t_3\in W$.
\end{example}

\begin{example}\label{eg:top}
	For a set \(V\), let \(\mathcal FV\) be the set of \emph{filters} on \(V\), upwards-closed subsets of \(\pow(V)-\{\emptyset\}\) that are closed under pairwise intersection.
	For a function \(f : V \to V'\), let \(\mathcal F(f)(F) = [f(F)]_{\text{fil}}\) be the smallest filter containing \(\{f(s) \mid s \in F\}\). 
	Then \(\mathcal F\) is an unbounded functor. 
	As Gumm points out in \cite{gumm2001functors}, the category \(\catfont{Top}\) of topological spaces and open continuous maps is a covariety of \(\mathcal F\)-coalgebras.
	The structure map of a topological space sends every point to its filter of neighbourhoods.
	The coequation presenting \(\catfont{Top}\) appears in \cite{kurz2005operations} in modal form, but in principle can be translated into a generalized coequation. 
\end{example}

\newcommand{\type}{\mathtt{T}}
\newcommand{\typeS}{\mathtt{S}}
\newcommand{\ato}{\mathtt{At}}
\section{Coequations-as-equations}\label{sec:coeq-eq}


Our main sources for this section are \cite{jacobs1995mongruences,hensel1994defining,cirstea1999coequational,rocsu2001equational}. 
The last two papers are written in the language of \emph{visible and hidden sorts}, making them relatively difficult to read.
Here, we follow the single-sorted setup of \cite{jacobs1995mongruences,hensel1994defining}. 
The generalisation to multiple sorts (\ie to the category $\Set^S$ for some set of sorts $S$) presents only notational difficulties.

Following \cite{jacobs1995mongruences}, let $\ato$ be a set of atomic types and consider the grammars of types:
\begin{equation*}
\typeS  ::= \mathtt{A}\in \ato \mid \mathtt{0}\mid \mathtt{1} \mid \typeS + \typeS \mid \typeS \times \typeS
 \qquad\qquad
\type  ::= \mathtt{A}\in \ato \mid \mathtt{0}\mid \mathtt{1} \mid \type + \type \mid \type \times \type \mid \mathtt{X}
\end{equation*}
A \emph{destructor signature} is a set of pairs of types $\sigma_i\defeq (\typeS_i,\type_i ), i\in I$, called \emph{destructors}.  
The interpretation of a type is determined inductively given interpretations \(\sem{\mathtt{A}}\) of $\mathtt{A}\in\ato$ and $\sem{\mathtt{X}}$ as sets, and by taking $+$ to be the coproduct, $\times$ the product, $0$ the empty set, and $1$ the set \(\{0\}\) in $\Set$.
An interpretation of a destructor $\sigma\defeq(\typeS,\type)$ is a map $\sem{\sigma}: \sem{\typeS}\times \sem{\mathtt{X}}\to \sem{\type}$, and an interpretation of a destructor signature is an interpretation of each of its destructors.

As $\Set$ is Cartesian closed, every destructor can equally be interpreted as a map $\sem{\sigma}: \sem{\mathtt{X}}\to \sem{\type}^{\sem{\typeS}}$. 
This means that the interpretation of a destructor signature can be described as a coalgebra for the functor
\[
T\sem{\mathtt{X}} \defeq \prod_{i\in I}\sem{\type_i}^{\sem{\typeS_i}}\qquad\qquad\text{($\type$ typically depends on $\mathtt{X}$). }
\]

\begin{example}\label{eg:bank}
Jacobs provides the example of a simple class for a bank account where $\ato=\{\mathtt{N}\}$, and the destructor signature is $\{(\mathtt{1},\mathtt{N}), (\mathtt{N}, \mathtt{X})\}$.  
An interpretation of this destructor signature can be defined by choosing $\sem{\mathtt{N}}=\N$ and two maps
\(
\mathrm{bal}: \sem{\mathtt{X}}\to \N
\)
and
\(
\mathrm{credit}: \N\times \sem{\mathtt{X}}\to \sem{\mathtt{X}}
\)
returning the balance on the account and crediting the account by a given amount respectively. Alternatively, the interpretation of this destructor signature can be a coalgebra for $\N\times \Id^{\N}$.
\end{example}
The definition of a \emph{term} for a destructor signature is fairly elastic (see \textit{op.cit}.) but includes at least the following rules. 
First, define for each atomic type $\mathtt{A}\in\ato$ a set $\mathrm{Var}_{\mathtt{A}}$ of variables of type $\mathtt{A}$. 
We also define a \emph{unique} variable $x$ of type $\mathtt{X}$. 
The following rules \cite{jacobs1995mongruences,cirstea1999coequational} are used to build terms in context:
\begin{enumerate}
\item Variables are terms of the corresponding type: if $a\in \mathrm{Var}_{\mathtt{A}}$ then $\vdash a:\mathtt{A}, \vdash x:\mathtt{X}$.
\item If $\sigma=(\typeS, \type)$ is in the destructor signature, then $s:\typeS,  t:\mathtt{X}\vdash \sigma(s,t): \type$.
\end{enumerate}
Any constructions which might be useful, such as projections and coprojections or built-in functions, can be added to the grammar of terms.
In the case of \cref{eg:bank} it is useful to add the function $(-)+(-): \mathtt{N\times N}\to\mathtt{N}$ 
which allows the term $n:\mathtt{N}\,x:\mathtt{X}\vdash \mathrm{bal}(x)+n:\mathtt{N}$ to be constructed.  
A \emph{coequation-as-equation} is defined as an equation $a_1: \mathtt{A}_1, \ldots,a_n:\mathtt{A}_n, x:\mathtt{X}\vdash s=t: \type$ between two terms of the same type, in the same context.
The \emph{interpretation of terms} follows in the obvious way from the interpretation of the destructor signature (and any other build-in operations like $(-)+(-)$) and function composition.

As in the case of \emph{coequations-as-corelations}, the purpose of these equations is \emph{not} to identify terms via a quotient, but to \emph{select} certain behaviours.  
The connection with corelations can be made explicit by observing that every term will be typed like $a_1: \mathtt{A}_1, \ldots,a_n:\mathtt{A}_n, x:\mathtt{X}\vdash t: \type$ with a context containing a \emph{unique} variable $x:\mathtt{X}$, and variables of atomic type. 
This means that its interpretation $\sem{t}$ can always be Curried, and since an interpretation of the destructor signature is a coalgebra $\gamma: X\to TX$, we can view a \emph{coequation-as-equation} $s=t$ as a corelation:
\[
\xymatrix@C=10ex
{
X \ar@<3pt>[r]^-{\sem{s}}\ar@<-1pt>[r]_-{\sem{t}} & \sem{\type}^{\sem{\mathtt{A}_1}\times \ldots\sem{\mathtt{A}_n}}
}
\]
or a pre-cocongruence
\[
\xymatrix@C=10ex
{
(X,\gamma) \ar@<3pt>[r]^-{\sem{s}}\ar@<-1pt>[r]_-{\sem{t}} & C_T\sem{\type}^{\sem{\mathtt{A}_1}\times \ldots\sem{\mathtt{A}_n}}
}
\]
Returning to \cref{eg:bank}, Jacobs gives the financially sound equation $n:\mathtt{N},x:\mathtt{X}\vdash\mathrm{bal}(\mathrm{credit}(n,x))=n+\mathrm{bal}(x)$. 
By interpreting the basic destructors as a coalgebra $\gamma: X\mapsto\N\times X^{\N}$, and Currying the interpretation of the two terms in the equation, we get a corelation $X\rightrightarrows \N^\N$ classifying behaviours according to what the functions $\lambda n.\sem{\mathrm{bal}(\mathrm{credit}(n,x))}$ and $\lambda n. n+\sem{\mathrm{bal}(x)}$ do at state $x$. 
The coequation-as-equation \emph{selects} the bank accounts whose behaviours cannot be distinguished by these two functions.

%
%
%

\section{Coequations-as-modal-formulas}\label{sec:modal}
Modal logic, and its generalisation coalgebraic modal logic, is an intuitive and powerful syntax to write \emph{predicate coequations}.
We follow the abstract formalism of \cite{kupke2004algebraic,kupke2005ultrafilter,jacobs2010exemplaric}, as it naturally lends itself to interpreting modal formulas as coequations. In this formalism, the syntax of a modal logic is given by an endofunctor $L$ on
some base category $\mathscr{C}$, typically either the category $\BA$ of Boolean Algebras for boolean modal logics (see \textit{op.cit.} and \cite{pattinson2003coalgebraic,venema2007algebrasandcoalgebras,cirstea2011modal}), or the category $\DL$ of Distributive Lattices (see \cite{balan2013positive,dahlqvist2015completeness,dahlqvist2017positivication}) for positive modal logics.  
The base category $\mathscr{C}$ encodes all the basic logical connectives whilst the functor $L$ constructs terms with modal operators. 
Since one is typically interested in finitary logics, we also make the natural assumption that $L$ is finitary, and in particular a varietor \cite[Thm. 3.17]{adamek2003varieties}.  We thus have a free-forgetful adjunction $F_L\dashv U_L, F_L: \mathscr{C}\to\mathsf{Alg}_{\mathscr{C}}(L)$.
We also assume that there exists a free-forgetful adjunction $\free\dashv \forg, \free: \mathscr{C}\to\Set$, which we write in sans-serif font to keep it distinct from the other adjunctions. 
\begin{example}\label{ex:modal}
	Normal modal logic is defined by the functor $L:\BA\to\BA$ which sends a boolean algebra $A$ to the boolean algebra of formal terms
	\(
	LA=\free\{\Box a\mid a\in \forg A\}/\equiv
	\),
	where $\equiv$ is the stable congruence generated by the equations $\Box\top = \top, \Box(a\wedge b)=\Box a\wedge\Box b$. 
	Usually, $A$ is also freely generated, specifically $A=\free \At$ where $\At$ is the set of propositional variables.
\end{example}
Coalgebraic modal logics are interpreted in coalgebras, so let us fix an endofunctor $T:\mathscr{D}\to\mathscr{D}$ describing both the kind of carrier ($\mathscr{D}$-objects) and the kind of transition systems in which modal formulas are to be interpreted.  
We now need two pieces of categorical data. 
\begin{enumerate}
\item A dual adjunction $G\dashv P, G:\mathscr{C}\to\mathscr{D}\op$  connecting the `logical category' $\mathscr{C}$ to the `model category' $\mathscr{D}$ whose objects carry the models of the interpretation. 
\item A \emph{semantic natural transformation} $\delta: LP\to PT$ to recursively compute the semantics.  
\end{enumerate}
\vspace{2pt}
The framework of coalgebraic modal logic can thus be summarized as the categorical data:
\begin{align*}
\xymatrix@C=7ex
{
\Set \ar@/^0.7pc/[rr]^-{\free} & {\scriptstyle\bot} & \mathscr{C} \ar@(ul, ur)[]^{L} \ar@/^0.7pc/[ll]^-{\forg} \ar@/^0.7pc/[rr]^-{G} & {\scriptstyle\bot} & \mathscr{D}\op  \ar@(ul, ur)[]^{T\op}  \ar@/^0.7pc/[ll]^-{P}  &  \delta:LP\to PT
}
\end{align*}
By using $P$ and $\delta$, one can turn any $T$-coalgebra $\gamma: X\to TX$ into an $L$-algebra $P\gamma\circ \delta_X: LPX\to PTX\to PX$. 
We denote this construction, which is functorial since $\delta$ is natural, by $\widehat{P}(X, \gamma)$.
The semantics can now be defined as follows: given a $T$-coalgebra $(X,\gamma)$ and a set $\At$ of variables, a \emph{valuation} is a map $v: \At\to \forg PX$, which lifts to a $\mathscr{C}$-morphism $\hat{v}: \free\At \to PX$. 
Since $PX$ is the carrier of $\widehat{P}(X,\gamma)$, we can re-type $\hat{v}$ as a $\mathscr{C}$-morphism $\hat{v}: \free\At \to U_L\widehat{P}(X,\gamma)$.
Since $L$ is a varietor, this map freely extends to a unique $L$-algebra morphism $\sem{-}_v:F_L\free\At\to  \widehat{P}(X,\gamma)$, which recursively computes the interpretation of a modal formula in $F_L\free\At$ as an element of $PX$ via the semantic transformation $\delta$.
A formula $\varphi\in F_L\free\At$ is said to be \emph{satisfied at $x\in X$ for the valuation $v$}, written $(X,\gamma,x)\models_v\phi$, if $x\in \sem{\phi}_v$. 
A formula $\varphi\in F_L\free\At$ is said to be \emph{valid in $(X,\gamma)$}, written $(X,\gamma)\models \varphi$, if it is satisfied at every $x\in X$ and for every valuation $v: \At\to \forg PX$.
\begin{example}\label{ex:modal2}
	In the case of the normal modal logic described in \cref{ex:modal} we take $\mathcal{D}=\Set$,  the dual adjunction is given by the powerset functor $\Pow: \Set\op\to\BA$ and the ultrafilter functor $\Uf: \BA\to\Set\op$,  and models are coalgebras for the powerset functor $\mathscr{P}:\Set\to\Set$.  
	The  semantic transformation $\delta: L\Pow\to\Pow \mathscr{P}$ thus turns a modal formula over a predicate on the carrier into the set of successors which must satisfy this predicate, from the perspective of the modality. 
	It is defined by
	\(
		\delta(\Box W) = \{W'\mid W'\subseteq W\}
	\).
	Given a coalgebra $\gamma: X\to\mathscr{P}X$ and a valuation $v: \At\to \forg\Pow X$, it follows from the definition of $\delta$ and $\sem{-}_v$ that for any $\varphi\in F_L\free\At$, $\sem{\Box \varphi}_v$ is computed recursively via $\sem{\Box \varphi}_v=\{x\in X: \gamma(x)\subseteq \sem{\varphi}_v\}$, where $\sem{p}_v\defeq v(p)$. This is the usual semantics for normal modal logic, rephrased coalgebraically.
\end{example}
A \emph{set} of modal axioms $\Phi\subseteq \forg U_L F_L\free\At$ defines a set of equations \[e_1,e_2: \Phi\rightrightarrows \forg U_LF_L \free\At,\] in the sense of \cref{subsec:alg}, since each axiom $\varphi\in\Phi$ is shorthand for the equation $\varphi=\top$, \ie $e_1(\varphi)=\varphi, e_2(\varphi)=\top$.  
We can then consider the variety defined by the coequalizer $q:F_L \free\At\epi Q$ of the free extensions $\hat{e}_1,\hat{e}_2: F_L \free \Phi\rightrightarrows F_L \free\At$, exactly as in \cref{subsec:alg}. 
We obtain, immediately from the definitions, that every set of modal axioms defines a \emph{variety}, and these axioms are valid in a coalgebra $(X,\gamma)$ precisely when $\widehat{P}(X,\gamma)$ belongs to the variety.

\begin{proposition}
Using the notation above,  $(X,\gamma)\models \Phi$ iff $q\perp \widehat{P}(X,\gamma)$.
\end{proposition}

A set of modal axioms $\Phi$ can also be seen as a coequation-as-predicate which defines a covariety.  To see this, assume that $T$ is a covarietor, \ie that there exists forgetful-cofree adjunction $U_T\dashv C_T, U_T: \mathsf{CoAlg}_{\mathscr{D}}(T)\to\mathscr{D}$, and consider the cofree coalgebra $C_T G\free\At$ over the $\mathscr{D}$-object of colours $G\free\At$. The reason for choosing these colours is that by using the counit $\varepsilon$ of the adjunction $U_T\dashv C_T$,   we can construct a canonical interpretation via the adjunctions $G\dashv P$ and $F_L\dashv U_L$, and the fact that $U_L\widehat{P}\simeq PU_T$:
\begin{align*}
	U_TC_TG\free\At \stackrel{\varepsilon}{\longrightarrow}G\free\At
	&\iff 
	\free\At\longrightarrow PU_TC_TG\free\At \\
	&\iff 
	F_L\free\At \stackrel{\sem{-}_{\varepsilon}}{\longrightarrow} \widehat{P}C_T G\free\At.
\end{align*}
With this canonical interpretation map we can view $\Phi$ as a coequation-as-predicate in $G\free\At$ colours selecting the elements $t\in C_T G\free\At$ such that $(C_T G\free\At,t)\models_{\varepsilon}\varphi$ for all $\varphi\in\Phi$.
\begin{example}
Returning to the classical modal logic of \cref{ex:modal,ex:modal2}, we modify the semantics slightly by considering coalgebras for an accessible version of $\mathscr{P}$, \eg we take $T\defeq\mathscr{P}_{\kappa}$ with $\kappa=\lvert\At\rvert$ so that $\mathscr{P}_\kappa\At=\mathscr{P}\At$.  The coalgebraic semantics is now given in terms of a covarietor.
The cofree coalgebra $C_T\mathcal{U}\free\At\simeq C_T\mathscr{P}\At$ is then the set of all $\kappa$-branching strongly-extensional trees labelled by sets of propositional variables~\cite{worrell1999terminal}. 
By definition of $\sem{-}_{\varepsilon}$,  if $p\in\At$ then $(C_T \mathscr{P}\At, t)\models_{\varepsilon} p$ iff $p$ belongs to the set of propositional variables labelling $t$. 
The semantics of modal formulas works in the expected way, namely $\Box\varphi$ holds at a tree $t$ if $\varphi$ holds at all its children (should it have any).
Thus, every modal formula $\varphi\in F_L\free\At$ defines the subset of $\sem{\varphi}_{\varepsilon}\subseteq U_TC_T\mathscr{P}\At$,  \ie a \emph{predicate coequation}.  
More generally, every set $\Phi$ of modal axioms defines the predicate coequation $\bigcap_{\varphi\in \Phi}\sem{\varphi}_{\varepsilon^T}\subseteq U_TC_T \mathscr{P}\At$, containing only those trees for which all formulas in $\Phi$ are satisfied. 
\end{example}

Now, which covarieties can be defined from a coequation-as-modal-formula in the way we just sketched? 
An answer to this question has long been part of the canon of modal logic, and is known as the Goldblatt-Thomason theorem \cite[Thm. 3.19]{blackburn2006handbook}. 
A coalgebraic version of this theorem was developed by Kurz and Rosick{\`y} in \cite{kurz2007goldblatt}.  
This theorem can only be stated for (and is therefore only applicable to) coalgebraic modal logics such that the semantic transformation $\delta: LP\to PT$ has an inverse natural transformation $\delta\inv: PT\to LP$\footnote{The naturality of the inverse in not strictly necessary, but makes the presentation easier, see \cite{kurz2007goldblatt}.}\textsuperscript{,}\footnote{The existence of a natural transformation $\delta\inv: GL\to TG$ is also key to the duality between varieties and covarieties and between equations and coequations developed in \cite{salamanca2016dualityofequations}. It is also crucial to strong completeness proofs in coalgebraic modal logic \cite{dahlqvist2015phd,dahlqvist2016coalgebraic}.}.
From this inverse we can construct a natural transformation $h: GL\to TG$, called its \emph{mate}  \cite{kurz2007goldblatt}. 
For our purposes, it is enough to say that $h$ and $G$ allow us to turn every $L$-algebra $\alpha: LA\to A$ into a $T$-coalgebra $h_A\circ G\alpha: GA\to GLA\to TGA$.  
We denote this operation $\widehat{G}(A,\alpha)$. 
In some sense, it is dual to the functor $\widehat{P}$ defined earlier. 
Given a $T$-coalgebra $(X,\gamma)$, its \emph{ultrafilter extension} is the $T$-coalgebra $\widehat{G}\widehat{P} (X,\gamma)$.  
A class $\mathsf{C}$ of $T$-coalgebras is \emph{closed under ultrafilter extensions} if $(X,\gamma)\in\mathsf{C}$ implies $\widehat{G}\widehat{P} (X,\gamma)\in \mathsf{C}$.   
A class $\mathsf{C}$ of $T$-coalgebras \emph{reflects ultrafilter extensions} if $\widehat{G}\widehat{P} (X,\gamma)\in \mathsf{C}$ implies $(X,\gamma)\in\mathsf{C}$. 

\begin{theorem}[Coalgebraic Goldblatt-Thomason Theorem \cite{kurz2007goldblatt}]
Let $T:\Set\to\Set$ preserve finite sets, and assume the existence of a natural inverse $\delta\inv: PT\to LP$ to the semantic transformation, then a class of $T$-coalgebras closed under ultrafilter extensions is definable by coequations-as-modal-formulas iff it is closed under homomorphic images, subcoalgebras, coproducts, and if it reflects ultrafilter extensions.
\end{theorem}


\section{Conclusion}\label{sec:prospects}




In this review we have presented four types of syntaxes for `writing a coequation': \emph{coequa\-tions-as-corelations}, which come with the special syntax of \emph{coequations-as-equations} for functors of the type $\prod_{i\in I} A_i\times \mathsf{P}_i^{B_i}$ where each $\mathsf{P}_i$ is polynomial, and \emph{coequations-as-predicates}, which come with the special syntax of \emph{coalgebraic modal logic}. It is worth emphasising that the corelation and the predicate perspective are semantically equivalent and one can move from one to the other by taking an equalizer or a cokernel pair respectively. However,  both the \emph{syntax} and the \emph{intuition} are different, and these aspects matter a great deal in practice. 

A rule of thumb for which syntax to use in which situation might be the following. Thinking of the elements of a cofree coalgebra as generalized trees, if the aim is to specify a behaviour defined by a relationship between a tree and (some of) its children, then the corelation perspective is probably the most useful.  This perspective was illustrated in \cref{sec:coeq-eq}, but can also be found in the beautiful work on stream differential equations of Hansen, Kupke and Rutten \cite{hansen2016stream}. On the other hand, if the aim is to enforce or avoid a particular \emph{behavioural pattern}, then the predicate perspective is probably the most useful.

Although much of our discussion focused on literature written decades ago, coequations continue to find new uses.
It was recently observed that coequations appear in formal language theory as \emph{varieties of languages}~\cite{ballesterbolinchescosmellopezrutten2015dual,SalamancaBBCR15}, which play a dual role to monoid equations.
A vastly wider perspective on this relationship was explored in subsequent work~\cite{Salamanca17,adamek2019generalized-eilenberg}.
For another example, a behavioural coequation appeared in a proof of the completeness of an axiomatisation of \emph{guarded Kleene algebra with tests} (\textsf{GKAT})~\cite{schmidkappekozensilva2021gkat}, an algebraic framework for reasoning about simple imperative programs. 
There, the coequation is the set of behaviours specified by terms in the expression language of \textsf{GKAT}, much like the coequation in \cref{eg:process algebra example}, and is used to present the covariety of automata that implement \textsf{GKAT} programs.
This usage of coequations may also be possible in the context of an open problem posed by Milner~\cite{milner1984complete}, as a covariety implicitly appears in a recent partial solution~\cite{grabmeyerfokkink2020complete}. 

As the examples above illustrate, the use value of coequations is emerging, slowly, from the literature.
Given the scope of their applications, we hope that our synthesis of the literature will make coequations more accessible to the general computer science community.

\subsubsection*{Acknowledgements}

The authors are most grateful to Alexander Kurz for his services as history consultant. The responsibility for any mistake or mischaracterisation lie solely with the authors.

\printbibliography

\end{document}